\documentclass[10pt, journal]{IEEEtran}
\usepackage{graphicx}
\usepackage{epsfig}
\usepackage{color}
\usepackage{caption}
\usepackage{subcaption}
\usepackage{multirow}
\usepackage{diagbox}
\usepackage{url}
\usepackage{algorithm}
\usepackage{algorithmic}
\usepackage{amssymb}
\usepackage{amsmath}
\usepackage{epstopdf}
\usepackage{latexsym}
\usepackage{wrapfig}
\usepackage[normalem]{ulem}
\usepackage[bottom]{footmisc}
\usepackage{footnote}
\usepackage{scrextend}
\usepackage{scrextend}
\usepackage{cite}

\usepackage{wasysym}
\usepackage{tikz}
\usepackage{xparse}

\NewDocumentCommand{\statcirc}{ O{#2} m }{%
    \begin{tikzpicture}
    \fill[#2] (0,0) circle (1.0ex); 
    \fill[#1] (0,0) -- (180:1ex) arc (180:0:1.5ex) -- cycle; 
    \end{tikzpicture}
}

\makesavenoteenv{tabular}
\makesavenoteenv{table}

\deffootnote[2.2em]{2.2em}{2.2em} {\textsuperscript{\thefootnotemark}}
\IEEEoverridecommandlockouts

\begin{document}


\title{A Survey on Consensus Protocols in Blockchain for IoT Networks}
 


\author{\IEEEauthorblockN{Mehrdad Salimitari and Mainak Chatterjee\\
\IEEEauthorblockA{Department of Computer Science\\
University of Central Florida\\
 Orlando, FL 32825\\
Email: \{mehrdad, mainak\}@cs.ucf.edu}
}
}

 \maketitle

\begin{abstract}
The success of blockchain as the underlying technology for cryptocurrencies has opened up 
possibilities for its use in other application domains as well. 
The main advantages of blockchain for its potential use in other  
domains are its inherent security mechanisms and immunity to different attacks. 
A blockchain relies on a consensus method for agreeing on any new data. 
Most of the consensus methods which are currently used for the blockchain of different cryptocurrencies require high computational power and thus are not apt for resource-constrained systems.

In this article, we discuss and survey the various blockchain based consensus methods  that are applicable to resource constrained IoT devices and networks. 
A typical IoT network consists of several devices which have limited computational and 
communications capabilities. Most often, these devices cannot perform intensive computations and are starved for bandwidth. 
Therefore, we discuss the possible measures that can be taken to reduce the computational power and convergence time for the underlying consensus methods. 
We also talk about some of the alternatives to the public blockchain like private blockchain
and tangle, along with their potential adoption for IoT networks.
Furthermore, we review the existing consensus methods that have been implemented and explore the possibility of utilizing them to realize a blockchain based IoT network. Some of the open research challenges are also put forward. 

\end{abstract}


\section{Introduction}

We are witnessing the proliferation of Internet of Thing (IoT) applications in our houses, offices, neighborhoods and cities. Adoption of IoT based technologies are poised to make a big impact on various sectors of our daily lives, including energy, manufacturing, intelligent transportation and smart cities. With more and more autonomous deployments of potentially large scale IoT systems, ensuring security, availability and confidentiality of the data, the devices and the networks become utmost critical~\cite{bhattacharjee2017preserving}.

In order to realize an entirely autonomous IoT network, different sensors and devices (generically referred to as `nodes') in an IoT network need to communicate with each other in a distributed fashion. This requires a mechanism by which different nodes in an IoT network can agree upon the validity of any communicated data. One of the best ways to achieve this is to use a consensus method. 
There exist different consensus methods by which different nodes can reach a common decision without the participation of a central controller~\cite{debus2017consensus}.
Consensus methods typically require high computational and communications capabilities. There are some that are less demanding, 
however, they do not provide strict guarantees on the performance attributes~\cite{zheng2016blockchain}.

As far as the computational and communication capabilities of the IoT devices are concerned, they are typically inexpensive low-powered embedded computing platforms
capable of performing their bare minimum tasks.
Most embedded IoT devices are equipped with 8-bit or 16-bit microcontrollers with very little RAM and storage capacities, and can connect to the Internet either 
via ethernet or low-powered wireless communications such as IEEE 802.15.4~\cite{sehgal2012management}. These resource constraints make it difficult to directly apply the traditional consensus protocols to IoT networks.
Newer technologies like blockchain~\cite{zheng2016blockchain} and tangle~\cite{popov2016tangle}, that use consensus methods for accepting new data, have the potential to be customized depending on the requirements
and constraints of various IoT devices and networks. 




This article reviews how blockchain works and argues that full-fledged implementations of blockchain 
(as in bitcoin and other cryptocurrencies) are impractical for resource-constrained IoT devices and networks. 
Though blockchain employs very compute-intensive hash operations for cryptocurrencies, such operations are not 
essential for non-critical systems as resource constrained systems might be willing to trade some level of data 
integrity for savings in computations and energy consumption.
Private blockchains, that allow participation of valid users and are not open to the public, 
have been proposed to overcome this problem. 
These blockchain implementations use consensus methods that do not require high computational power for solving hash problems.
However, since they are private, only valid users can access them. We compare the pros and cons of various implementations
and discuss their applicability towards IoT networks. We highlight some of the research challenges that need to be 
addressed for widespread deployment of blockchain based IoT networks. 

 The rest of the article is organized as follows. 
 In section~\ref{sec-2}, we briefly discuss what blockchain is. 
 In section~\ref{IOT}, we discuss what the requirements of IoT networks and also what the
 limitations of the current technologies are. 
 In section~\ref{BC}, we motivate how blockchain can be applied for IoT networks
 with appropriate modifications. 
 In section~\ref{sec-3}, we discuss different consensus methods.
 In section~\ref{sec-implementations}, we survey the existing implementations of blockchain and discuss their pros and cons. In section~\ref{research-section}, we put forward some of the open research challenges. Conclusions are drawn in the last section. 

\vspace{-2mm}
\section{An overview of Blockchain}
\label{sec-2}

A blockchain is a distributed and tamper-resistant database that no single entity controls, 
but can be shared and accessed by all. 
New records (called blocks) can be added to the existing blocks as long as the new block is approved by all in the network. 
Also, once blocks are recorded, it is not feasible to modify or erase them. 
Blockchains have been designed to work on an unreliable network with adversarial entities. 
By using sophisticated and compute-intensive secure hash algorithms, blockchain achieves data integrity which prevents data erasure or manipulation and invalid information from being recorded. These compute-intensive algorithms are part of 
a {\em proof of work} which is a method of consensus by which different nodes in a network can agree upon new data or detect an anomaly. 
However, there exist other consensus methods with significantly lower computational requirement and network overhead that we will review in section~\ref{consensus}.

Proof of work is a computationally expensive consensus approach. In this method, different nodes in a blockchain try to solve a cryptographic hash function, SHA-256 in particular. SHA-256 generates a unique, fixed size 256-bit hash.
The process of finding the next block through solving this hash problem is called {\em mining}
and the users performing this task are called {\em miners}. 
The solved block contains transactions, hash of the previous block, nonce and a time stamp. 
After a miner finds a valid nonce value which solves this problem, he releases his solved block so other miners can be aware of it. 
In the next step, other miners verify the claimed block by using the block information and its claimed associated nonce as an input to a SHA-256 function. If its output is less than the target value, the miner will accept it as a valid block and withdraw all of his effort for solving that block and move on to find the next block~\cite{salimitari2017profit}. The miners' incentive for verifying a block is to avoid spending time and computational resources on an already solved block. On the other hand, it is vital for miners to be the one who finds the next block while validating a new block. What withholds miners from accepting a new block without validating it and  mining the next block is that whenever the blockchain detects an invalid block, it will discard that block and all the blocks built upon it. A block gets more credibility when more blocks are built upon it.
This is known as confirmation in cryptocurrencies' blockchains. After several confirmations, a block and its contained transactions are considered to be acceptable.

Blockchain stores all data transactions in chained blocks, i.e., a block contains several transactions 
linked to previous blocks by a hash pointer~\cite{nakamoto2008bitcoin}. 
This chain continues to the first block mined in blockchain which in bitcoin blockchain was mined by \textit{Satoshi Nakamoto} and is called the \textit{Genesis block}. 
This architecture makes it infeasible to modify the transactions in the blockchain because in order to change a transaction, it is required to change all the blocks built on that transaction in less than 10 minutes. 
This modification might be feasible with more than half of the hash rate power of the world.
In other words, in order to add a block to the blockchain, the miner should find a specific \textit{nonce} value for each block in a way that the hash value of that block be less than a specific target value. This process is mathematically hard and time consuming because it can only be done by brute force search and the miners have to try different nonce values randomly to reach the target value~\cite{salimitari2017profit}. 
However, adversaries can affect blockchain operation of cryptocurrencies such as bitcoin by gaining control over 25\% of computation power via an attack known as selfish mining~\cite{Eyal:2018}. 
Although this attack cannot jeopardize the immutability of bitcoin's blockchain, it can increase revenue of adversaries.

\vspace{-2mm}
\subsection{An Illustrative Example}

\begin{figure} [!h]
\begin{center}
\includegraphics[width=3.5in, height=1.5in,trim=0.15cm 0 0 0cm]{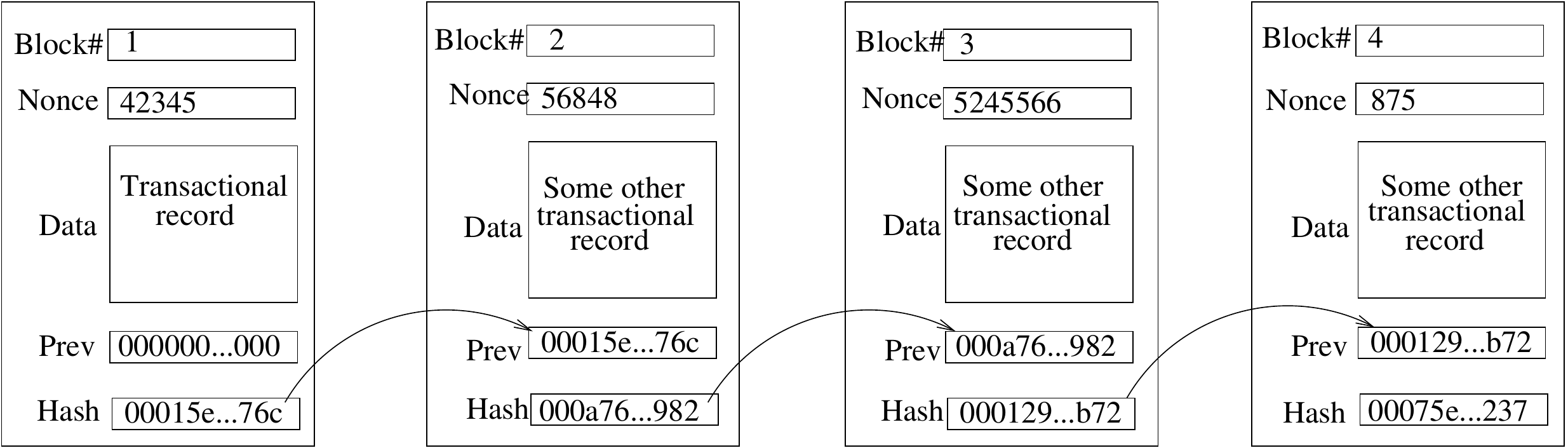}
\caption{Chaining of blocks in blockchain}
\label{fig-blockchain}
\end{center}
\vspace{-4mm}
\end{figure}

Figure~\ref{fig-blockchain} illustrates a blockchain of 4 blocks each of which contains the {\tt Block number}, a (random number) {\tt Nonce}, 
the {\tt data}, the hash of the previous block (i.e., {\tt Prev}) and 
the hash of the current block (i.e., {\tt Hash}). 
Since there is no block prior to block 1, the previous hash is set to all 0's. With the block number, data and previous hash known, the objective is to find the nonce such that the hash of all is less than a target value, i.e., 
$\mbox{SHA-256 }( {\tt  Block \#, Nonce, Data, Prev) \le target ~ value}.$ 
Suppose the nonce found is 42345 as shown in block 1. 
The target value indicates the difficulty level of the proof of work. 
In other words, this target value mandates that the 256-bit output of the hash function starts with a certain number of
consecutive zeros. In the illustration, all hashes are shown to start with three 0's. 
For block 2, the same process is repeated to obtain the nonce with the new block number and the data. 
Also, {\tt Prev} is updated with the {\tt Hash} of the previous block. Again, we observe that a different nonce is obtained and the {\tt Hash} 
begins with three consecutive 0's. 
It is to be noted, as the target value decreases, the difficulty of finding the nonce increases. 

\vspace{-6mm}
\subsection{Private Blockchain}
\label{private blockchain}
There exist some blockchains designed by different companies for specific applications with restricted access to the public. The most known private blockchains are part of the Hyperledger project which is a collaboration between many well-known companies and hosted by the Linux foundation. Having access to these blockchains or participating in their consensus protocol is permissioned and dependent on a third-party. Therefore, private blockchains are centralized to some extent which is in contradiction to the original idea of blockchain being decentralization and entirely distributed. 
However, these blockchains have lower computational requirements and faster network response time 
which make them more desirable for IoT applications. 
In some private blockchains, some of the nodes have complete access to all the stored blocks in the blockchains. Thus, private blockchains bring more privacy to the information which is desirable for companies. 

\begin{figure}[!h]
 \begin{center}
   \subcaptionbox{Public Blockchain}{
     \includegraphics[width=2.1in,height=1.7in]{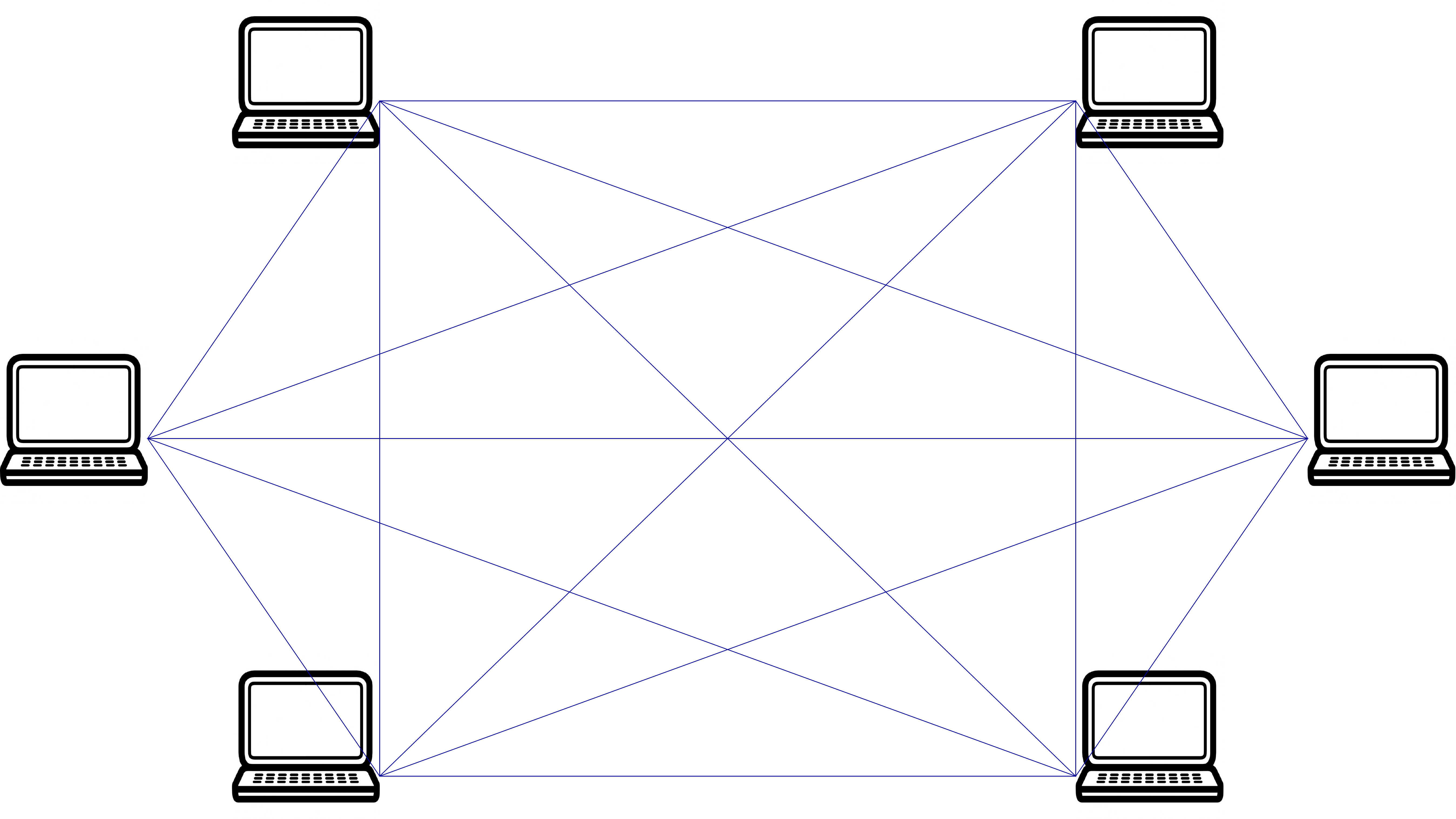}\hspace*{0em}
     \label{public}
   }\qquad \qquad \qquad
   \subcaptionbox{Private Blockchain}{
     \hspace*{0em}\includegraphics[width=2.1in,height=1.7in]{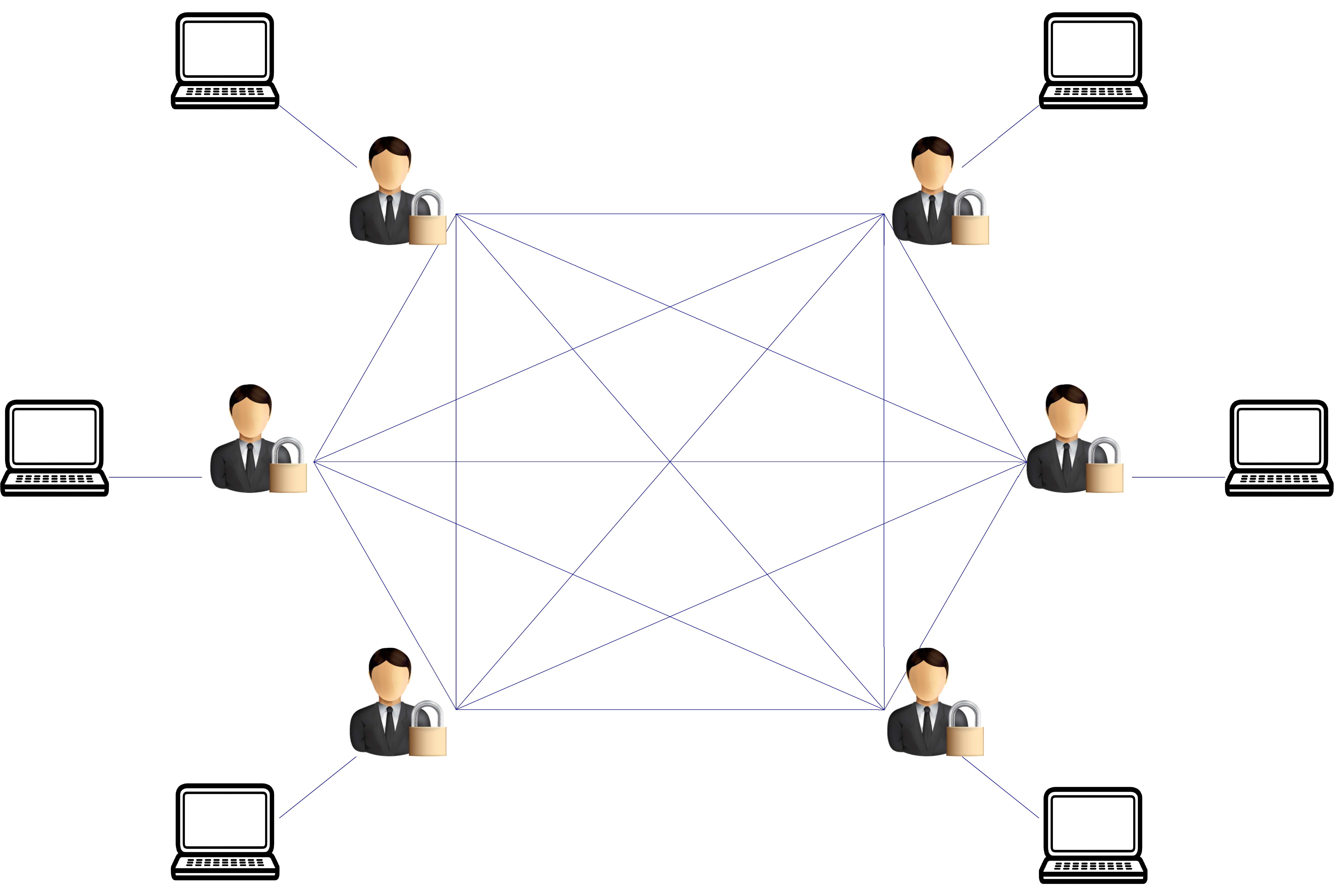}
     \label{private}
   }
   \caption{{\small Public blockchain vs private blockchain. In a private blockchain, users' access to the blockchain is restricted by an institution.}}
  \label{public vs private}
 \end{center}
 \end{figure}

Private blockchains are more secure than traditional databases because of using cryptographic protocols similar to public blockchains. However, they are not as secure as public blockchains that employ computational-intensive protocols like proof of work. Thus, there is the possibility of tampering stored data in private blockchains. 
It should be mentioned that there exist some partially private blockchains called consortium blockchains. While a fully private blockchain is controlled by a single company, consortium blockchains are governed by several institutions 
all of which directly participate in the consensus protocol.
 Private blockchains can follow different consensus methods like practical Byzantine fault tolerance, proof of elapsed time and proof of stake
which have been discussed in section~\ref{consensus}. Private and public blockchains are compared in Fig.~\ref{public vs private}.

\vspace{-6mm}
\section{IoT: Importance and Limitations}
\label{IOT}
We are witnessing the use of IoT networks in various domestic, industrial and military applications. 
A common feature of these IoT networks is that they 
consist of several sensors and actuators which are resource-constrained devices capable of communication without human intervention. 
Besides these devices, there are other network entities that connect the sensors and actuators to the backbone network infrastructure.
These are routers, switches, aggregators and cloud infrastructure comprising virtual servers and storage-- all of which
dictate the baseline requirements for resource provisioning and sharing.
These requirements include dynamic and verifiable group membership of devices, authentication and data integrity, robustness against single point of failure, lightweight operations in terms of resources and low latency communication~\cite{banafa2017iot}.

\vspace{-2mm}
\subsection{Smart Homes: An example of an IoT Network}
One of the first uses of IoTs has been the automation of homes and cities which are referred as \textit{smart homes} and \textit{smart cities}. A smart home involves creating a local network consisting of various smart devices, sensors, and meters within a house. This network is made accessible from anywhere to the residents of the house through the Internet. 
One of the main obstacles towards realizing this goal is securing the communicated and stored data in this network against malicious acts that desire to wreak havoc with someone's home.

In  a smart home network, there is a significant amount of data that is exchanged among different nodes. 
Electricity meters need to access all the electrical appliances in the house. 
It has to interact with the the smart car for charging and discharging (during peak times, one might use 
the stored energy in the car as a backup for electricity demands). 
The HVAC system needs to communicate with the thermostat and weather forecasting systems. 
Similarly, the fire extinguisher receives signals from smoke detectors. 
All these connections, including the residents need to remotely access the devices, 
require exchange if data in a secure manner. 

In addition, in a smart home, there is a need to store all the communicated data among the devices
and the current state of each device and sensor for analysis and better decision making. 
For example, historical data along with resident's power consumption habits can be used for for energy saving 
decisions and recommendations. 
Also, some smart homes can learn and adapt
the device parameters based on historical observations. In such cases, system logs are very instrumental in accurate learning, model building, predictive analysis, and decision making~\cite{wang2015processing}. 
Needless to emphasize, the stored data needs to be immune to data manipulation attacks. 

Extending the concept of smart homes, one can imagine of a {\em smart city} as large network connecting various smart homes, smart transportation systems, smart power grids, smart vehicles, and so on. Securing such a large network is a bigger concern since any 
vulnerability in any of the smaller network can be exploited to launch bigger attacks with significant damages
dues to cascading effects. Compared to a smart home, a smart city's requirements are more stringent
and demands even a more secure communication links and data storage/exchange protocols. 

Though there are many competing technologies that try to immune communication links and stored data in smart homes against attacks, 
blockchain has emerged as probably the most promising one. 
In a blockchain, the data is immutable because of the underlying consensus protocols.
Therefore, manipulation attacks on transmitted or stored data are not plausible through a single compromised node and majority of nodes ought to be compromised for a successful attack~\cite{salimitari2017profit}. The suitability of blockchain for IoT networks is discussed in details in Section~\ref{BC}.

\subsection{Current IoT Technologies and Limitations}
Most of the existing IoT solutions rely on the centralized server-client paradigm which are connected to cloud servers via the Internet. Although these solutions are suitable for the present day, considering the growth of IoT networks, there is a demand for new solutions to make the network more decentralized~\cite{fernandez2018review}. One of the suggested solutions is the creation of large peer-to-peer (P2P) wireless sensor networks. 
However, this solution does not adequately address the privacy and security requirements for IoT networks~\cite{triantafillou2003nanopeer},\cite{ali2004csn},\cite{krco2005p2p}. 
Following are some of the main challenges for existing IoT implementations that need to be
to be adopted for large-scale deployments.
\begin{itemize}
\item {\em High cost:} Most of the current IoT solutions require centralized clouds and server farms which impose high costs for deployment and maintenance. Also, these requirements necessitate middlemen which would make the infrastructure more centralized and costly.

\item {\em Maintenance:} Apart from the high cost related to maintenance of millions of smart devices within the network, applying required software updates to all of these devices would be a considerable burden.

\item {\em Privacy:} For extensive deployment of IoT networks, users need to be assured about the privacy and anonymity of their data. Some companies permit certain authorities (i.e., governments, manufacturers or service providers) to access and control users' devices. 

\item {\em Security:} Existing IoT solutions use closed-source codes which do not provide transparency to users. Future IoT implementations would require utilization of open-source codes so that all the users can maintain and 
update the software making it less susceptible to malicious activities. In addition, current IoT implementations are prone to single point of failure. If a cloud server confronts a physical or software malfunction, it may impact the entire network functionality~\cite{kshetri2017can}. Also, a breach in a single device connected to a server or the cloud can wreak havoc with the entire system through denial of service (DoS) attacks by sending malicious messages to other devices, leaking private data, or manipulating the gathered data~\cite{anirudh2017use},\cite{xu2016security},\cite{li2017iot},\cite{yu2017recursive}. 
Collected data from different devices is stored, processed and forwarded by intermediate systems which can tamper the data. In addition, wireless channels used for data broadcast are unreliable which makes the network prone to jamming attacks~\cite{wang2019survey}. More detailed analysis of possible security breaches in IoT networks are examined in~\cite{lin2017survey} and~\cite{khan2018iot}.

\end{itemize}

On the other hand, all blockchain implementation have a decentralized network which eliminate some of the current IoT limitations like high cost and maintenance. In addition, the inherent consensus methods used in different blockchain implementations can address privacy and security issues in IoT networks which are surveyed in section~\ref{consensus}.

\section{Blockchains for IoT Networks}
\label{BC}
Given the stringent requirements for IoT networks, blockchain appears to be very apt for both 
i) securing the network against manipulation attacks that target stored data and 
ii) providing a secure platform for all devices in the network to communicate with each other~\cite{lin2017using, novo2018blockchain}. 
Let us discuss the shortcomings of the current models and what promises do blockchain bring to
the IoT paradigm.


\subsection{Client-Server vs. P2P Models}

Most of the currently used IoT networks are based on the server-client model where all the devices are identified, authenticated and connected via cloud servers requiring enormous amount of processing capability and storage capacity. In addition, all the communications between these devices have to go through the Internet even if the devices are close to each other. Though such a model is practical for small IoT networks, it does not scale well. Furthermore, the cost of establishing large number of communication links, maintaining centralized clouds and networking all equipment, is significant for large-scale IoT networks. Apart from the costs, reliance on cloud servers make the architecture susceptible to single point of failure. Moreover, IoT devices must be immune to information attacks and physical tampering. Though some of the existing methods make the IoT devices secure, they are complex and not appropriate for resource-constrained IoT devices with limited computation power~\cite{banafa2017iot}.

\subsection{Blockchain for Resource Constrained IoT Networks}

Blockchain establishes a peer-to-peer network which decreases the cost of installation and maintenance of centralized clouds, data centers and networking equipment by distributing the computational and storage requirements among all the devices within the network. This communication paradigm solves the single point of failure problem. Blockchain addresses the privacy concerns for IoT networks by using cryptographic algorithms. It also solves the reliability issues in IoT networks by using tamper-resistant ledgers~\cite{banafa2017iot}.

In spite of the built-in mechanisms that guarantee data integrity in blockchain based systems, the
implementation of it in resource constrained IoT networks is challenging
due to the following reasons. First, computation of the cryptographic hashes as part of the consensus method is
compute-intensive and demands immense CPU cycles. Second, the communication links could become a
bottleneck in delivering the transactions to others and getting their approvals. The problem is even more
aggravated in an interference-limited wireless system when all IoT devices have to vie for shared radio links. Third, IoT networks consist of many devices that need to communicate with each other very quickly and at all times. 
This necessitates adding many blocks containing large number of transactions to the blockchain every 
second which requires low latency consensus methods.

Typically, resource constrained IoT systems have some flexibility in their performance requirements 
and ready to trade some level of  data integrity for savings in computations and energy consumption. 
One of the ways to achieve that is to relax the proof of work in order to reduce computations requirements. Another
way could be not to maintain {\em all} the blocks since as the number of blocks increase, larger storage is required.
Rather, the last $l$ blocks could be chained and all computations could be performed on the last $l$ blocks. 
Though such techniques would not yield  the level of data integrity usually 
provided by cryptocurrencies, they would nevertheless guarantee some reduced 
level of data protection. 
Other ways to empower blockchain to be used for resource-constrained IoT networks is to use consensus methods
that have significantly lower computational requirements, network overhead and faster convergence. 
These methods are discussed next.

\section{Consensus Methods}
\label{sec-3}
\label{consensus}

There are several well-established methods by which different nodes in a blockchain network can reach consensus over a new block. 
A blockchain based system is as secure and robust as its underlying consensus method.
The most well-known consensus method is proof of work (discussed in Section~\ref{sec-2}) which is used by bitcoin. Proof of work has proved to be an effective approach for cryptocurrencies over years. However, due to its high computational and bandwidth requirements, it does not seem to be practical for IoT networks. 
Therefore, we present other existing consensus methods and discuss the possibility of applying them to a blockchain based IoT network. Then, we compare all the discussed consensus protocols in Table~\ref{table-consensus}.
The consensus protocols that are not applicable to IoT networks or are mere modifications of a general consensus method 
are discussed in brief.

\subsection{Proof of Work (PoW)}
There are some consensus methods that based on proof of work.
They are discussed next.

\subsubsection{Proof of Capacity (PoC)}
Proof of Capacity  is similar to PoW but instead of depending on the computing power of the miners, it relies on their hard disk capacity. Thus, it is significantly more energy-efficient than ASICs mining used in PoW. In PoC, miners have to store huge data sets, known as plots, to get an opportunity to mine the next block. Therefore, by storing more plots, a miner will have a higher probability to solve the next block~\cite{debus2017consensus}. 
The block creation time in PoC is 4 minutes. 
PermaCoin and SpaceMint are two cryptocurrencies employing PoC. 
 Aside from the high latency, this method is not a rational choice for IoT networks where the devices have limited storage capacity. 
\vspace{2mm}
\subsubsection{Proof of Elapsed Time (PoET)}
Proof of Elapsed Time is a consensus method proposed by Intel which works similar to PoW but with significantly lower energy consumption. In this method, miners have to solve a hash problem similar to that of PoW. However, instead of a competition between miners to solve the next block, the winning miner is randomly chosen based on a random wait time. The winning miner is the one whose timer expires first. The verification of correctness of timer execution is done using a Trusted Execution Environment (TEE) like Intel's Software Guard Extension (SGX)~\cite{hyperledger}.
 
PoET's eased computational requirements make it IoT friendly. In addition, its low latency and high throughput make it favorable for IoT networks. 
The main drawback of this approach is its dependency on Intel which is in conflict with the basic philosophy of blockchain being entirely decentralized. 

\vspace{-1mm}
\subsection{Proof of Stake (PoS)}
This is the most prevalent method of consensus in the blockchains used for cryptocurrencies after proof of work. This method works similar to proof of work but with a significant difference. It does not induce in a race amongst the nodes  to solve the next block. Instead of a competition between the nodes to solve the next block, a node is chosen by lottery to solve the next block. The node that will mine the next block is chosen based on its proportional stake in the network which is its wealth in terms of that cryptocurrency. The chosen node will use a digital signature to prove its ownership over the stake instead of solving a complicated hash problem. As a result, this method does not require high computational power. In this method, all the coins (i.e., the cryptocurrency) are available from the first day and no mining reward or coin creation exists and the miners are rewarded only with a transaction fee~\cite{debus2017consensus}.

 Although this method eliminates the computational requirements of proof of work, it creates new problems. This method is contingent upon nodes with the highest amount of stake which somehow makes the blockchain centralized. Furthermore, there is another problem called ``nothing at stake" which refers to the situation in which a selected node has nothing to lose if it behaves badly. Therefore, nothing prevents a node from, for example, creating two sets of new blocks to obtain more reward for transaction fees. However, some modifications are applied to this method to deal with these problems. Although, this method has significantly eased computational requirements of proof of work, it is not yet popular 
for resource constrained IoT networks. 
In addition, this method and its variants  (which are discussed below) are based on monetary concepts (stakes) that does not exist in IoT networks.
\vspace{2mm}

\subsubsection{Delegated Proof of Stake (DPoS)}
This method is based on the proof of stake consensus method. Contrary to PoS which is direct democratic, this method is representative democratic~\cite{zheng2016blockchain}
which means that all the stakeholders vote to choose some nodes as \textit {witnesses} and \textit{delegates}. Witnesses are responsible and rewarded for creating new blocks. The delegates are responsible for maintaining the network and proposing changes such as block sizes, transaction fees, or reward amount. 
In each election round, $N$ witnesses with the highest votes are chosen. $N$ is defined such that at least 50\% of the voting stakeholders believe there is enough decentralization. It should be noted that the number of witnesses that each stakeholder vote for,
must be at least equal to their desired number of witnesses for decentralization. 
In DPoS, there are built-in mechanisms in place for detecting and voting out a malicious delegate or witness~\cite{debus2017consensus, bitshare, larimer2014delegated}. 

Bitshares, a cryptocurrency,  uses DPoS which has shown to significantly improve throughput and latency compared to  PoS, but at the cost of making the blockchain more centralized. It is capable of processing 100,000 transactions per second (TPS). In addition, a block is added to the blockchain in 1.5 seconds on average and in 3 seconds  maximum~\cite{debus2017consensus, bitshare, larimer2014delegated}. 
Though these features make DPoS very attractive for IoT networks, 
the main bottleneck for DPoS in IoT networks is its dependency on monetary concepts (stakes) to choose witnesses and delegates.

\vspace{2mm}
\subsubsection{Leased Proof of Stake (LPoS)}
Leased Proof of Stake works the same way as PoS but with some improvements. LPoS tries to solve the centrality problem in PoS. It enables the nodes with low balances to participate in block verification by adding a leasing option. Leasing allows the wealth holders with higher balances to lease their funds for specific amount of time to nodes with low balances. The leased amount will be in possession of the wealth holders during the lease contract, however, it will increase the chance of solving a block for nodes with low balances. When these nodes solve a block, they will share the reward with the wealth holders proportionally. This approach makes blockchain more secure by making it more decentralized~\cite{lease}. 
Unfortunately, LPoS is not useful in the context of IoT since it is based on monetary concepts which is not necessarily applicable to an IoT network.

\vspace{2mm}
\subsubsection{Proof of Importance (PoI)}
Proof of Importance is a modified version of PoS where instead of considering only nodes' balances to determine the next winning node for solving the next block, it takes into account more factors including a node's reputation which is specified by a particular system defined function and the number of transactions occurred to or from that node. 
Therefore, this method of consensus considers productive network activity of nodes which is more efficient than only nodes' balances~\cite{importance}. The cryptocurrency NEM
uses PoI for consensus.
PoI yields  high throughout and exhibits comparatively low latency. 
In addition, it does not require large computational or network overheads. 
These features make PoI favorable for IoT networks. However, PoI like other PoS-based methods, is also dependent on monetary concepts. 

\vspace{2mm}
\subsubsection{Proof of Activity (PoA)}
Proof of Activity is a hybrid method of consensus based on proof of work and proof of stake~\cite{zheng2016blockchain}. First, miners try to solve a hash function in a competition to find the next block as in proof of work. However, the solved block will only contain a header and the miner's address without any transaction. Then, transactions are added to the block and according to the solved block's header, a group of validators is chosen to sign the new block in order to reach consensus. This step is done by using proof of stake. This approach is safer against attacks but can experience higher delay which might not be acceptable for delay-sensitive IoT applications~\cite{debus2017consensus}.
\vspace{2mm}
\subsubsection{Casper}
 Casper is a PoS-based consensus protocol for transitioning Ethereum's consensus method from PoW to PoS. It is an overlay on the existing PoW blockchain in Ethereum~\cite{buterin2017casper}. It prevents ``nothing at stake" attacks by slashing all the attacker's stakes. A modification of the {\em Greedy Heaviest Observed Subtree (GHOST)} is used as the chain selection rule in Casper. GHOST is proposed as an alternative for the longest-chain rule used in the bitcoin blockchain. In blockchain networks, there is a possibility for forks because of the network delay. In those cases, the longest chain is selected as the main chain and other branches are ignored. 
 In the GHOST protocol, at each fork, the heaviest subtree rooted at the fork is selected in the chain~\cite{sompolinsky2015secure,zamfir2017casper}. This method provides significant benefits for cryptocurrencies in terms of security and delay. 
 However, they are not sufficient for tackling the IoT challenges in comparison with the naive PoS method.

\vspace{2mm}
\subsubsection{Proof of Burn (PoB)}
Proof of Burn  is based on burning coins which refers to sending coins to an irretrievable address. Miners get priority to solve the next block according to the amount of coins they have burnt~\cite{debus2017consensus}. While this approach is practical for designing cryptocurrencies, it is not appropriate for IoT applications 
since this method is contingent upon existence of a monetary framework and burning of coins,
neither of which is inherent in an IoT networks. 
A cryptocurrency called Slimcoin uses PoB.

\subsection{Byzantine Agreement Methods}
\begin{figure} [h]
 \begin{center}
 \includegraphics[width=3.6in, height=2.7in,trim=0cm 0 0cm 0cm]{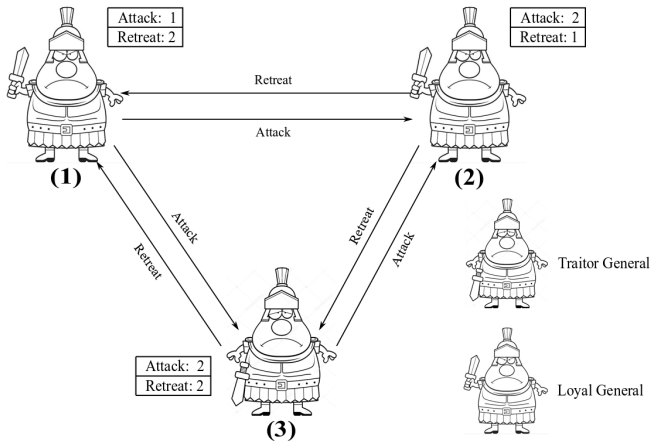}
 \caption{{\small Byzantine problem with 3 generals. PBFT can only tolerate malicious activity by less than 1/3 of nodes. 
 In this example, general 3 is traitor who can prevent the loyal generals to a reach consensus in case they have different opinions by sending them a different decision from their own decisions. Here, loyal general 1 decides to attack and loyal general 2 decides to retreat. Traitor general 3 uses the arisen opportunity from loyal generals' conflict of opinions and sends them two different messages containing a decision in contrast with their own decisions. Therefore, based on the received messages, general 1 decides to retreat and general 2 decides to attack.}}
 \label{pbft}
 \end{center}
 \vspace{-6mm}
 \end{figure}

These consensus methods are based on Byzantine generals problem. In short, this problem is to come up with a consensus method between Byzantine generals over the attack strategy with the assumption that some generals may be traitors and try to adopt treacherous actions to prevent loyal generals reach a consensus and conquer the city. The scenario for this problem is illustrated in Fig.~\ref{pbft}. 
The followings are the Byzantine Agreement-based consensus methods.

\vspace{2mm}
\subsubsection{Practical Byzantine Fault Tolerance (PBFT)}
In this method, all the nodes should participate in the voting process in order to add the next block and the consensus is reached when more than two-thirds of all nodes agree upon that block. PBFT can tolerate malicious behavior from up to one-third of all nodes to perform normally. For instance, in a system with one malicious node, there should be at least 4 nodes to reach a correct consensus. Otherwise, consensus  is not reached.
In this method, the consensus is reached quicker and more economically compared to proof of work.
Also, it does not require owning assets similar to proof of stake to take part in the consensus process~\cite{debus2017consensus}. 

This method is well-suited for private blockchains like the Hyperledger projects that are controlled by a third-party. However, it is not the best choice for permissionless, public blockchains due to their limited scalability and comparatively low tolerance towards malicious activities. PBFT has high throughput, low latency, and low computational overhead-- all of which are desirable for IoT networks. However, its high network overhead makes it un-scalable for large networks, thus it could be applied only to small IoT networks.

 \vspace{2mm}
\subsubsection{Delegated Byzantine Fault Tolerance (dBFT)}
Delegated Byzantine Fault Tolerance follows the same rules as PBFT but it does not require participation of all the nodes for adding a block which make it more scalable. In dBFT, some nodes are chosen as delegates of other nodes and according to some rules, they 
pursue the consensus protocol similar to PBFT~\cite{zheng2016blockchain}. A cryptocurrency called NEO uses this method of consensus. 
This method has many desirable features similar to that of PBFT. However, its average latency for block creation is 15 seconds which is not acceptable for an IoT network.
\vspace{2mm}
\subsubsection{Stellar Consensus Protocol (SCP)}
Stellar Consensus Protocol provides micro-finance services on the blockchain platform. 
It was proposed by Mazieres using a variant of PBFT called federated byzantine fault tolerance (FBFT) as the backbone~\cite{mazieres2015stellar}. In FBFT, nodes belonging to intersecting groups (i.e., the federates) run a local consensus protocol among their members~\cite{dinh2018untangling}.
This method is decentralized and is open to the public which allows everyone to participate in the consensus protocol. It has a very low latency similar to web transactions (a few seconds at most). This is the first Byzantine agreement based consensus method which provides users with the maximum freedom to choose among different combinations of other participants to trust in order to reach a consensus. 

SCP reaches robustness through quorum slices. A quorum is a set of nodes that participate in the consensus protocol and a quorum slice is its subset that helps a node in its agreement process. Individual trust decisions made by participants in the blockchain constructs a quorum slice, and quorum slices connect the whole network together in a similar fashion as peering networks create the Internet by binding together. It should be noted that the quorums are selected by the nodes involved in the transactions. 

SCP consists of two steps: nomination protocol and ballot protocol. First, nomination protocol is executed. During this step, new values called candidate values are proposed for agreement. These values are sent to all the nodes in the quorum and each of them will vote for a single value among the candidate values. At the end of process, unanimously values are chosen for that slot. Then, ballot protocol is initiated which involves a federated voting to either accept or abort the obtained values in the nomination protocol. Finally, the ballot for the current slot is finalized and aborted ballots are discarded. In situations that nodes cannot reach a consensus to abort or to commit a value, a higher valued ballot is initiated which can be considered as a new ballot protocol execution~\cite{mazieres2015stellar},~\cite{sankar2017survey}.

High throughput and low computational requirements of this method are desirable for IoT networks. Although its latency is comparatively low, it is not in order of milliseconds which is required for IoT networks. If the latency of this method could be decreased by reducing the network overhead (which is $O(n)$), this method can be a good choice for IoT networks.

\vspace{2mm}
\subsubsection{Ripple}
Similar to Stellar, Ripple uses FBFT consensus method. It is proposed to reduce the latency of blockchains~\cite{schwartz2014ripple}. In this decentralized method, each miner uses a trusted subset of nodes within the larger network to reach a consensus. There are two types of nodes in the network: server nodes which are responsible for the consensus protocol and client nodes which only transfer funds. Each server node contains an Unique Node List (UNL). Nodes within an UNL are used for reaching consensus over new transactions. When 80\% of the nodes within an UNL agree over a transaction, the consensus is reached. 

The Ripple consensus protocol is executed every few seconds by all the nodes in order to reach consensus over new transactions. Ripple can tolerate up to 20\% faulty nodes in the UNL~\cite{zheng2017overview}. This method is mostly used for monetary purposes to enable transactions with no chargebacks~\cite{chalaemwongwan2018state}. 
It has similar features as Stellar which makes it a good candidate for IoT networks if its latency is decreased to order of milliseconds.

\vspace{2mm}
\subsubsection{Tendermint}
Tendermint belongs to a family of Byzantine Fault Tolerance (BFT) consensus protocols which can host arbitrary application states~\cite{dib2018consortium}. It is a permissioned consensus method~\cite{zheng2017overview}. Contrary to PBFT where each node has equal voting power, in Tendermint nodes have different voting powers that are proportional to their stakes\cite{dinh2018untangling}. Therefore, it can be considered as a hybrid consensus method based on PBFT and PoS. Tendermint can tolerate malicious activity from at most one-third of the total byzantine voting power~\cite{Tendermint}. 

Participants in the Tendermint protocol are called validators which propose blocks of transactions and vote on them in turn. The voting process consists of two steps: pre-vote and pre-commit. A block is added to the blockchain when more than two-thirds of validators pre-commit for the same block in the same round~\cite{kwon2014tendermint}. Unlike PBFT, validators are chosen based on PoS by locking their coins and dishonest validators are punished~\cite{bach2018comparative}. Thus, this method relies on monetary concepts like PoS which is not necessarily applicable to IoT networks. Considering high scalability, high throughput, and low latency, this method could be moulded for IoT networks if the monetary concept can be replaced by some other criteria.

\vspace{2mm}
\subsubsection{ByzCoin}
ByzCoin is a byzantine consensus that uses a collective signing protocol called CoSi~\cite{syta2016keeping} to commit bitcoin transactions within seconds and makes PBFT scalable. It also modifies PBFT to enable a public blockchain by supporting dynamic membership proportional to PoW as in bitcoin. This method uses a tree-structured communication protocol which significantly reduces the latency of bitcoin's blockchain. 
However, it is vulnerable to DoS attacks~\cite{kogias2016enhancing}. 

This method yields a throughput higher than the Paypal network with a confirmation latency of 15-20 seconds~\cite{kogias2016enhancing}. Although  scalability and throughput of this method is desirable for IoT networks, its latency is beyond the acceptable limits of IoT networks. Furthermore, the PoW used in this method is similar to that of bitcoin~\cite{kogias2016enhancing} which is again not appropriate for IoT devices.

\subsection{VRF-based methods}
In verifiable random function (VRF)-based methods, 
committee members are randomly selected for participation in the consensus protocol~\cite{micali1999verifiable}. 
Algorand and Dfinity are two such methods.

\vspace{2mm}
\subsubsection{Algorand}
Algorand is a novel public and permissionless blockchain implementation that uses pure proof of stake (PPoS) consensus protocol built on byzantine agreement~\cite{Algorand}. 
Algorand has been proposed as a new cryptocurrency to address limitations of existing implementations: decentralization, scalability, and security. Most of the available blockchain implementations are centralized to some degree; i.e., in a bitcoin network, the users with the highest hash power control the network. However, in Algorand, each block is approved by a unique committee of users that are randomly selected by VRFs in a private and non-interactive fashion using the users' private key and public information from the blockchain. Thus, it is fully decentralized. This selection is based on users' weights assigned according to the amount of money in their account. This step is based on proof of stake~\cite{gilad2017algorand}. 

After the committee members are selected, the committee reaches a consensus over the new block using a byzantine agreement protocol called $BA\star$. Existing byzantine fault tolerance protocols need a fixed set of servers to be determined ahead of time which makes them vulnerable to Sybil attacks, where an adversary creates many pseudonyms to affect the Byzantine agreement protocol. In addition, they do not scale to a large number of users. However, $BA\star$ eliminates the risk of Sybil attacks and can scale to millions of users because of the existing first step in Algorand. $BA\star$ can tolerate up to 1/3 of weighted users (1/3 of the money owned by users) to be malicious like any other BFT based methods. However, because of using randomly selected VRFs in the first step, adversaries do not know which users to attack until they start participating in $BA\star$ which significantly increases the security of the consensus protocol~\cite{gilad2017algorand}. 

The main drawback of Algorand is that the randomness used in producing the VRF seeds can be biased by an adversary. To overcome this shortcoming, it uses a look-back mechanism to ensure strong synchrony and unbiased seeds. However, this makes the network vulnerable to  ``nothing at stake" attack~\cite{zamani2018rapidchain}.

Latency of Algorand is less than a minute  with transaction finality of around one minute. This feature is very desirable for a cryptocurrency. For instance, in bitcoin network it takes about 10 minutes to grow a chain by one block and it requires to wait for 6 blocks to ensure transaction finality (transaction be on an authoritative chain) which takes about an hour. There is not also a possibility of fork in Algorand~\cite{Algorand, gilad2017algorand}. 

Although Algorand propounds a bright horizon for financial purposes, it lacks several features to be successfully applied to IoT networks. Algorand does not require communication among users to determine if they are selected to participate in the consensus protocol. It also does not postulate computational resources to solve cryptographic puzzles. These characteristics alongside with decentralization, high scalability, and high security are appealing features of Algorand for IoT. However, an IoT network requires a delay less than a second (in order of milliseconds). Furthermore, there is no monetary concept in an IoT network to define stakes and assign weights.

\vspace{2mm}
\subsubsection{Dfinity}
Dfinity blockchain proposes a four-layer consensus protocol which can be used for permissioned networks or paired with Sybil resistance methods (e.g., proof-of-work or proof-of-stake) for permissionless and public networks. The most important feature of Dfinity is containing decentralized randomness beacon acting as the VRFs to produce random outputs as the seed. This new VRF protocol is based on a unique-deterministic, non-interactive, DKG-friendly threshold signature scheme which solves the biased coin problem and ``nothing at stake" problem which exist in Algorand. Dfinity consensus is also immune to selfish mining attacks~\cite{hanke2018dfinity}.  Dfinity can reach consensus over a new block in seconds and transaction finality after two confirmations~\cite{hanke2018dfinity}. Similar to Algorand, this high latency is not acceptable for IoT applications.

\subsection{Sharding-based methods}
Sharding is splitting the overheads of processing transactions among multiple,
smaller groups of nodes called shards~\cite{zamani2018rapidchain}. These methods were initially proposed to address the scalability problems in blockchain. Shards work in parallel to maximize the performance of the blockchain (processing more transactions in each consensus round) by sharding different overheads of blockchain network including communication, computation, and storage overhead. The stat-of-the-art sharding methods are discussed below.
\vspace{2mm}
\subsubsection{RSCoin}
RSCoin is a cryptocurrency framework proposed for centrally-banked cryptocurrencies to maintain complete control over the monetary supply using distributed set of authorities called mintettes (validator nodes). It uses a sharding-based approach to make traditional banking systems (with centralized monetary supply) more transparent via a distributed network while maintaining the scalability of the network. Danezis and Meiklejohn claim that their framework avoids double-spending~\cite{danezis2015centrally}. However, its two-phase commit protocol is not Byzantine fault tolerant and is susceptible to double-spending attacks by a colluding adversary~\cite{zamani2018rapidchain}. Moreover, this framework is not decentralized as it is asserted since it relies on a trusted source of randomness for sharding.

Although this framework has a very low latency (in the order of milliseconds) and high throughput~\cite{danezis2015centrally}, it is not applicable for IoT networks. It is mostly centralized and based on a centralized monetary supply which has no place in an IoT network.

\vspace{2mm}
\subsubsection{Elastico}
This is the first sharding-based consensus protocol proposed for public and permissionless blockchains that uses a combination of classical byzantine consensus protocol (e.g., PBFT) and proof of work to improve bitcoin blockchain. It partitions the network into smaller committees which process disjoint set of transactions (shards). The number of committees grows near linearly in the total computational power of the network. Each committee contains a small number of nodes that reach a consensus over their set of transactions (shard) using classical byzantine consensus. In each consensus round, these committees are selected using PoW's least-significant bits which are obtained by each node solving a PoW puzzle based on the consensus epoch randomness obtained from the last state of the blockchain~\cite{luu2016secure}.

Although this method can enhance latency and throughput of the bitcoin blockchain, it confronts several limitations: (i) It still suffers from a large communication overhead which is troublesome for IoT networks. (ii) The randomness used during each consensus epoch can be biased by an adversary which compromises the committee selection process. (iii) While each node is responsible for verifying a subset of transactions (a shard), it still has to broadcast all blocks to other nodes and store the entire ledger which does not address communication overhead and resource-constrained IoT devices. (iv) It can only tolerate up to 1/4 faulty nodes~\cite{zamani2018rapidchain}. 

Unfortunately, Elastico does not seem practical for IoT implementations for several reasons: (i) While improving bitcoin's latency, the latency  is still large enough for IoT networks. 
(ii) It has  very large communication overhead. (iii)
The storage requirements are not minimalist.
(iv) There are still security issues.

\vspace{2mm}
\subsubsection{OmniLedger}
OmniLedger is a permissionless and public blockchain implementation that uses VRFs and sharding with a ByzCoin-based consensus protocol called ByzCoinX. It has been proposed by Kokoris-Kogias et. al. to address some of the limitations of Elastico, specially latency and scalability. ByzCoinX improves the PBFT part of ByzCoin in order to increase its tolerance to DoS attacks. During each consensus epoch, the committee selection process is done by VRFs in a similar fashion as the lottery algorithm of Algorand. These committees reach a consensus over a subset of transaction (a shard). Similar to Elastico, it can tolerate only up to 1/4 faulty nodes. When there are less than 1/8 faulty nodes, this method can reach a comparatively low latency (less than 10 seconds)~\cite{kokoris2018omniledger}. 

Aside from the undesirable latency of this protocol for IoT networks and its security issues, this method poses a large communication overhead which cannot be implemented on resource-constrained IoT devices. In OmniLedger's consensus protocol, each committee has to communicate with all the nodes for each block of transaction. In addition, there is another communication overhead during producing the random seed for the VRF~\cite{zamani2018rapidchain}. However, OmniLedger provides low storage overhead since validators are not required to store the full transaction history which is in favor of IoT devices~\cite{kokoris2018omniledger}.
\vspace{2mm}
\subsubsection{RapidChain}
RapidChain has been proposed to solve existing limitations of sharding-based consensus methods including communication overhead, security, scalability and latency. Unlike other sharding-based methods which have limited tolerance to Byzantine faults, it can tolerate up to 1/3 faulty nodes without any trusted setup. RapidChain is the first public blockchain that provides full sharding: communication, computation, and storage overhead. This method significantly enhances latency and throughput by eliminating the need for linear amount of communication per transaction. This protocol does not require gossiping transactions to the entire network which improves the communication overhead~\cite{zamani2018rapidchain}.

The full sharding used in this method is very desirable for resource constrained IoT devices to successfully participate in the consensus protocol. In addition, RapidChain has a high throughput (processing 7,380 transaction per second in a network of 4,000 nodes). However, its latency for confirming the transactions is not suitable for IoT networks (8.7 seconds for 4,000 nodes). If the latency of this method can be improved, it would be a worthy  option for IoT networks~\cite{zamani2018rapidchain}.

\vspace{-5mm}
\subsection{Raft}
Raft is  a voting-based consensus method that was proposed to make Paxos algorithm more understandable and implementable for practical systems. Paxos algorithm tries to solve the consistency problem in certain conditions for Byzantine Generals Problem. Raft achieves the same efficiency as Paxos~\cite{mingxiao2017review, dib2018consortium}. Raft and Paxos are non-Byzantine fault tolerance algorithms. While their protocols are similar to BFT algorithms, they can only tolerate crash faults up to 50\% of the nodes while Byzantine algorithms can tolerate arbitrary (including maliciously) corrupted nodes. It is composed of two stages: leader election and log replication. The leader is responsible for ordering the transactions. The leader selection stage is executed using a randomized timeout for each server when an existing leader fails. When a leader is chosen, log replication stage is triggered. In this stage, the leader accepts log entries from clients and broadcasts transactions to make its version of the transaction log~\cite{dib2018consortium, ongaro2014search}. 
Corda and Quorum are two blockchain implementations that use Raft as their consensus method~\cite{dinh2018untangling}.

This method has high throughput and low latency. However, its throughput and performance depend on the leader node which occupies an absolute dominance in the system. Therefore, if the leader node is maliciously infected, the entire system will be destroyed. It cannot tolerate malicious nodes and can endure 50\% nodes of crash
fault~\cite{mingxiao2017review}. 
Since it is crucial to secure the leader node,  the
throughput is limited by the performance of that node. 
Due to its low security and restricted throughput, it is not very appropriate
for IoT networks. 

\begin{table*}[!h]
\scriptsize
\caption{Comparisons of different consensus methods~\cite{mingxiao2017review, xu2017taxonomy, chalaemwongwan2018state, bach2018comparative, zheng2017overview, kokoris2018omniledger, zamani2018rapidchain, bano2017consensus}}.
\centering

\begin{tabular}{|c|c|c|c|c|c|c|c|c|c|c|}
\hline
\textbf{Consensus} & \multirow{2}{*} {\textbf{Accessibility\footnotemark}} & \textbf{Decentral-} & \multirow{2}{*} {\textbf{Scalability}} & \textbf{Through-\footnotemark} & \multirow{2}{*} {\textbf{Latency\footnotemark}} & \textbf{Adversary} & \textbf{Computing} & \textbf{Network} & \textbf{Storage} & \textbf{IoT}\\
\textbf{method} &  & \textbf{ization} & & \textbf{put} & & \textbf{tolerance}  & \textbf{overhead} &  \textbf{overhead} & \textbf{overhead} & \textbf{suitability}\\
\hline
\multirow{2}{*} {\textit{\textbf{PoW}}} & Public, & \multirow{2}{*} {High} & \multirow{2}{*} {High} & \multirow{2}{*} {Low} & \multirow{2}{*} {High} & \textless25\% & \multirow{2}{*} {High} & \multirow{2}{*} {Low} & \multirow{2}{*} {High} & \multirow{2}{*} {\tikz\draw[black,fill=white] (0,0) circle (0.2cm);} \\
 & PL. & &  &  &  & Computing Power & & & &\\
\hline
\multirow{2}{*} {\textit{\textbf{PoC}}} & Public, & \multirow{2}{*} {High} & \multirow{2}{*} {High} & \multirow{2}{*} {Low
} & \multirow{2}{*} {High
} & \multirow{2}{*} {N/A} & \multirow{2}{*} {Low} & \multirow{2}{*} {Low} & \multirow{2}{*} {Very High} & \multirow{2}{*} {\tikz\draw[black,fill=white] (0,0) circle (0.2cm);} \\
 & PL. & &  &  &  & & & & &\\
\hline
\multirow{2}{*} {\textit{\textbf{PoET}}} & Private, & \multirow{2}{*} {Medium} & \multirow{2}{*} {High} & \multirow{2}{*} {High} & \multirow{2}{*} {Low} & \multirow{2}{*} {N/A} & \multirow{2}{*} {Low} & \multirow{2}{*} {Low} & \multirow{2}{*} {High} & \multirow{2}{*} {\tikz\draw[black,fill=black] (0,0) circle (0.2cm);}\\
& P. or PL. &  &  &  &  &  &  &   &  & \\
\hline
\multirow{2}{*} {\textit{\textbf{PoS}}} & Public & \multirow{2}{*} {High} & \multirow{2}{*} {High} & \multirow{2}{*} {Low} & \multirow{2}{*} {Medium} & \textless51\% & \multirow{2}{*} {Medium} & \multirow{2}{*} {Low} & \multirow{2}{*} {High} & \multirow{2}{*} {\begin{tikzpicture}
\draw (0,0) circle (0.2cm);
\clip (0,0) circle (0.2cm);
\fill[black] (0cm,0.2cm) rectangle (-0.2cm,-0.2cm);
\end{tikzpicture}} \\
& P. or PL. &  &  &  &  & Stakes & & & &\\
\hline
\multirow{2}{*} {\textit{\textbf{DPoS}}} & Public, & \multirow{2}{*} {Medium} & \multirow{2}{*} {High} & \multirow{2}{*} {High} & \multirow{2}{*} {Medium} & \textless51\% & \multirow{2}{*} {Medium} & \multirow{2}{*} {N/A} & \multirow{2}{*} {High} & \multirow{2}{*} {\begin{tikzpicture}
\draw (0,0) circle (0.2cm);
\clip (0,0) circle (0.2cm);
\fill[black] (0cm,0.2cm) rectangle (-0.2cm,-0.2cm);
\end{tikzpicture}}  \\
 & PL. & &  &  &  & Validators & & & &\\
\hline
\multirow{2}{*} {\textit{\textbf{LPoS}}} & Public & \multirow{2}{*} {High} & \multirow{2}{*} {High} & \multirow{2}{*} {Low} & \multirow{2}{*} {Medium} & \textless51\% & \multirow{2}{*} {Medium} & \multirow{2}{*} {Low} & \multirow{2}{*} {High} & \multirow{2}{*} {\tikz\draw[black,fill=white] (0,0) circle (0.2cm);}\\
& PL. &  &  &  &  & Stakes & & & &\\
\hline
\multirow{2}{*} {\textit{\textbf{PoI}}} & Public & \multirow{2}{*} {High} & \multirow{2}{*} {High} & \multirow{2}{*} {High} & \multirow{2}{*} {Medium} & \textless51\% & \multirow{2}{*} {low} & \multirow{2}{*} {Low} & \multirow{2}{*} {High} & \multirow{2}{*} {\begin{tikzpicture}
\draw (0,0) circle (0.2cm);
\clip (0,0) circle (0.2cm);
\fill[black] (0cm,0.2cm) rectangle (-0.2cm,-0.2cm);
\end{tikzpicture}} \\
& PL. &  &  &  &  & importance & & & &\\
\hline
\multirow{2}{*} {\textit{\textbf{PoA}}} & Public, & \multirow{2}{*} {High} & \multirow{2}{*} {High} & \multirow{2}{*} {Low} & \multirow{2}{*} {Medium} & \textless51\% & \multirow{2}{*} {High} & \multirow{2}{*} {Low} & \multirow{2}{*} {High} & \multirow{2}{*} {\tikz\draw[black,fill=white] (0,0) circle (0.2cm);} \\
 & PL. & &  &  &  & Online Stakes & & & &\\
\hline
\multirow{2}{*} {\textit{\textbf{Casper}}} & Public, & \multirow{2}{*} {High} & \multirow{2}{*} {High} & \multirow{2}{*} {Medium} & \multirow{2}{*} {Medium} & \textless51\%  & \multirow{2}{*} {Medium} & \multirow{2}{*} {Low} & \multirow{2}{*} {High} & \multirow{2}{*} {\tikz\draw[black,fill=white] (0,0) circle (0.2cm);} \\
 & PL. & &  &  &  & Validators & & & &\\
\hline
\multirow{2}{*} {\textit{\textbf{PoB}}} & Public, & \multirow{2}{*} {High} & \multirow{2}{*} {High} & \multirow{2}{*} {Low} & \multirow{2}{*} {High} & \textless25\%  & \multirow{2}{*} {Medium} & \multirow{2}{*} {Low} & \multirow{2}{*} {High} & \multirow{2}{*} {\tikz\draw[black,fill=white] (0,0) circle (0.2cm);} \\
 & PL. & &  &  &  & Computing Power & & & &\\
\hline
\multirow{2}{*} {\textit{\textbf{PBFT}}} & Private, & \multirow{2}{*}{Medium} & \multirow{2}{*}{Low} & \multirow{2}{*}{High} & \multirow{2}{*}{Low} & \textless33\% & \multirow{2}{*}{Low} & \multirow{2}{*}{High} & \multirow{2}{*}{High} & \multirow{2}{*} {\tikz\draw[black,fill=black] (0,0) circle (0.2cm);}\\
 & P. &   &  &  &  & Faulty Replicas & &  &  &\\
\hline
\multirow{2}{*} {\textit{\textbf{dPBFT}}} & Private, & \multirow{2}{*}{Medium} & \multirow{2}{*}{High} & \multirow{2}{*}{High} & \multirow{2}{*}{Medium
} & \textless33\% & \multirow{2}{*}{Low} & \multirow{2}{*}{High} & \multirow{2}{*}{High} & \multirow{2}{*} {\begin{tikzpicture}
\draw (0,0) circle (0.2cm);
\clip (0,0) circle (0.2cm);
\fill[black] (0cm,0.2cm) rectangle (-0.2cm,-0.2cm);
\end{tikzpicture}} \\
 & P. &   &  &  &  & Faulty Replicas & &  &  &\\
\hline
\multirow{2}{*} {\textit{\textbf{Stellar}}} & Public, & \multirow{2}{*} {High} & \multirow{2}{*} {High} & \multirow{2}{*} {High
} & \multirow{2}{*} {Medium} & \multirow{2}{*} {Variable}& \multirow{2}{*} {Low} & \multirow{2}{*} {Medium
} & \multirow{2}{*} {High} & \multirow{2}{*} {\begin{tikzpicture}
\draw (0,0) circle (0.2cm);
\clip (0,0) circle (0.2cm);
\fill[black] (0cm,0.2cm) rectangle (-0.2cm,-0.2cm);
\end{tikzpicture}}\\
& PL. &  &  &  &  & & &  &  &\\
\hline
\multirow{2}{*} {\textit{\textbf{Ripple}}} & Public & \multirow{2}{*} {High} & \multirow{2}{*} {High} & \multirow{2}{*} {High} & \multirow{2}{*} {Medium} & \textless20\% & \multirow{2}{*} {Low} & \multirow{2}{*} {Medium} & \multirow{2}{*} {High} & \multirow{2}{*} {\begin{tikzpicture}
\draw (0,0) circle (0.2cm);
\clip (0,0) circle (0.2cm);
\fill[black] (0cm,0.2cm) rectangle (-0.2cm,-0.2cm);
\end{tikzpicture}}\\
& PL. &  &  &  &  & Faulty UNL nodes &  &  &  & \\
\hline
\multirow{2}{*} {\textit{\textbf{Tendermint}}} & Private, & \multirow{2}{*} {Medium} & \multirow{2}{*} {High} & \multirow{2}{*} {High} & \multirow{2}{*} {Low} & \textless33\% & \multirow{2}{*} {Low} & \multirow{2}{*} {High
} & \multirow{2}{*} {High} & \multirow{2}{*} {\begin{tikzpicture}
\draw (0,0) circle (0.2cm);
\clip (0,0) circle (0.2cm);
\fill[black] (0cm,0.2cm) rectangle (-0.2cm,-0.2cm);
\end{tikzpicture}} \\
& P. &  &  &  &  & Voting power & &  &  & \\
\hline
\multirow{2}{*} {\textit{\textbf{ByzCoin}}} & Public, & \multirow{2}{*} {High} & \multirow{2}{*} {high} & \multirow{2}{*} {high} & \multirow{2}{*} {Medium} & \textless33\% & \multirow{2}{*} {high} & \multirow{2}{*} {Medium} & \multirow{2}{*} {high} & \multirow{2}{*} {\tikz\draw[black,fill=white] (0,0) circle (0.2cm);} \\
& PL. &  &  &  &  & Faulty Replicas & & &  &\\
\hline
\multirow{2}{*} {\textit{\textbf{Algorand}}} & Public, & \multirow{2}{*} {High} & \multirow{2}{*} {High} & \multirow{2}{*} {Medium} & \multirow{2}{*} {Medium} & \textless33\% & \multirow{2}{*} {Low} & \multirow{2}{*} {High} & \multirow{2}{*} {High} & \multirow{2}{*} {\tikz\draw[black,fill=white] (0,0) circle (0.2cm);} \\
& PL. &  &  &  &  & Weighted Users & & & & \\
\hline
\multirow{2}{*} {\textit{\textbf{Dfinity}}} & Public, & \multirow{2}{*} {High} & \multirow{2}{*} {High} & \multirow{2}{*} {N/A} & \multirow{2}{*} {Medium} & \multirow{2}{*} {N/A} & \multirow{2}{*} {Low} & \multirow{2}{*} {N/A} & \multirow{2}{*} {N/A} & \multirow{2}{*} {\tikz\draw[black,fill=white] (0,0) circle (0.2cm);} \\
 & P. or PL. & &  &  &  &  & & & &\\
\hline
\multirow{2}{*} {\textit{\textbf{RSCoin}}} & Private, & \multirow{2}{*} {Low} & \multirow{2}{*} {High} & \multirow{2}{*} {High} & \multirow{2}{*} {Low} & \multirow{2}{*} {N/A} & \multirow{2}{*} {Low} & \multirow{2}{*} {Medium} & \multirow{2}{*} {High} & \multirow{2}{*} {\tikz\draw[black,fill=white] (0,0) circle (0.2cm);}\\
& P. &  &  &  &  &  & & &  & \\
\hline
\multirow{2}{*} {\textit{\textbf{Elastico}}} & Public, & \multirow{2}{*} {High} & \multirow{2}{*} {High} & \multirow{2}{*} {Low
} & \multirow{2}{*} {High
} & \textless25\% & \multirow{2}{*} {Medium} & \multirow{2}{*} {High} & \multirow{2}{*} {High} & \multirow{2}{*} {\tikz\draw[black,fill=white] (0,0) circle (0.2cm);}\\
& PL. &  &  &  &  & Faulty VAlidators & &  &  & \\
\hline
\multirow{2}{*} {\textit{\textbf{OmniLedger}}} & Public, & \multirow{2}{*} {High} & \multirow{2}{*} {High} & \multirow{2}{*} {High} & \multirow{2}{*} {Medium} & \textless25\% & \multirow{2}{*} {Medium} & \multirow{2}{*} {Medium
} & \multirow{2}{*} {Low} & \multirow{2}{*} {\begin{tikzpicture}
\draw (0,0) circle (0.2cm);
\clip (0,0) circle (0.2cm);
\fill[black] (0cm,0.2cm) rectangle (-0.2cm,-0.2cm);
\end{tikzpicture}} \\
& PL. &  &  &  &  & Faulty Validators &  &  &  & \\
\hline
\multirow{2}{*} {\textit{\textbf{RapidChain}}} & Public, & \multirow{2}{*} {High} & \multirow{2}{*} {High} & \multirow{2}{*} {High} &\multirow{2}{*} {Medium} & \textless33\% & \multirow{2}{*} {Medium} & \multirow{2}{*} {Low} & \multirow{2}{*} {Low} & \multirow{2}{*} {\begin{tikzpicture}
\draw (0,0) circle (0.2cm);
\clip (0,0) circle (0.2cm);
\fill[black] (0cm,0.2cm) rectangle (-0.2cm,-0.2cm);
\end{tikzpicture}} \\
& PL. &  & &  &  & Faulty Validators &  &  &  & \\
\hline
\multirow{2}{*} {\textit{\textbf{Raft}}} & Private, & \multirow{2}{*}{Medium} & \multirow{2}{*}{High} & \multirow{2}{*}{High} & \multirow{2}{*}{Low} & \textless50\%  & \multirow{2}{*}{Low} & \multirow{2}{*}{N/A} & \multirow{2}{*}{High} & \multirow{2}{*} {\begin{tikzpicture}
\draw (0,0) circle (0.2cm);
\clip (0,0) circle (0.2cm);
\fill[black] (0cm,0.2cm) rectangle (-0.2cm,-0.2cm);
\end{tikzpicture}}\\
& P. &  &  &  &  & crash fault &  &  &  & \\
\hline
\multirow{2}{*} {\textit{\textbf{Tangle}}} & Public, & \multirow{2}{*}{Medium} & \multirow{2}{*}{High} & \multirow{2}{*}{High} & \multirow{2}{*}{Low} & \textless33\% & \multirow{2}{*}{Low} & \multirow{2}{*}{Low} & \multirow{2}{*}{Low} & \multirow{2}{*} {\tikz\draw[black,fill=black] (0,0) circle (0.2cm);} \\
& PL. &  &  &  &  & Computing Power &  &  &  & \\
\hline
\end{tabular}
\label{table-consensus}
\end{table*}

\footnotetext[1]{PL. denotes permissionless and P. denotes Permissioned.}
\footnotetext[2]{Low throughput is less than 100 TPS (transaction per second), medium throughput is between 100 TPS and 1000 TPS, and high throughput is more than 1000 TPS.}
\footnotetext[2]{High latency is in order of minutes, medium latency is in order of seconds, and low latency is on order of milliseconds.}

\subsection{Tangle}
\label{sec-tangle}
\begin{figure}[h]
\vspace{2mm}
  \centering
\includegraphics[width=3.5in, height=1.7in]{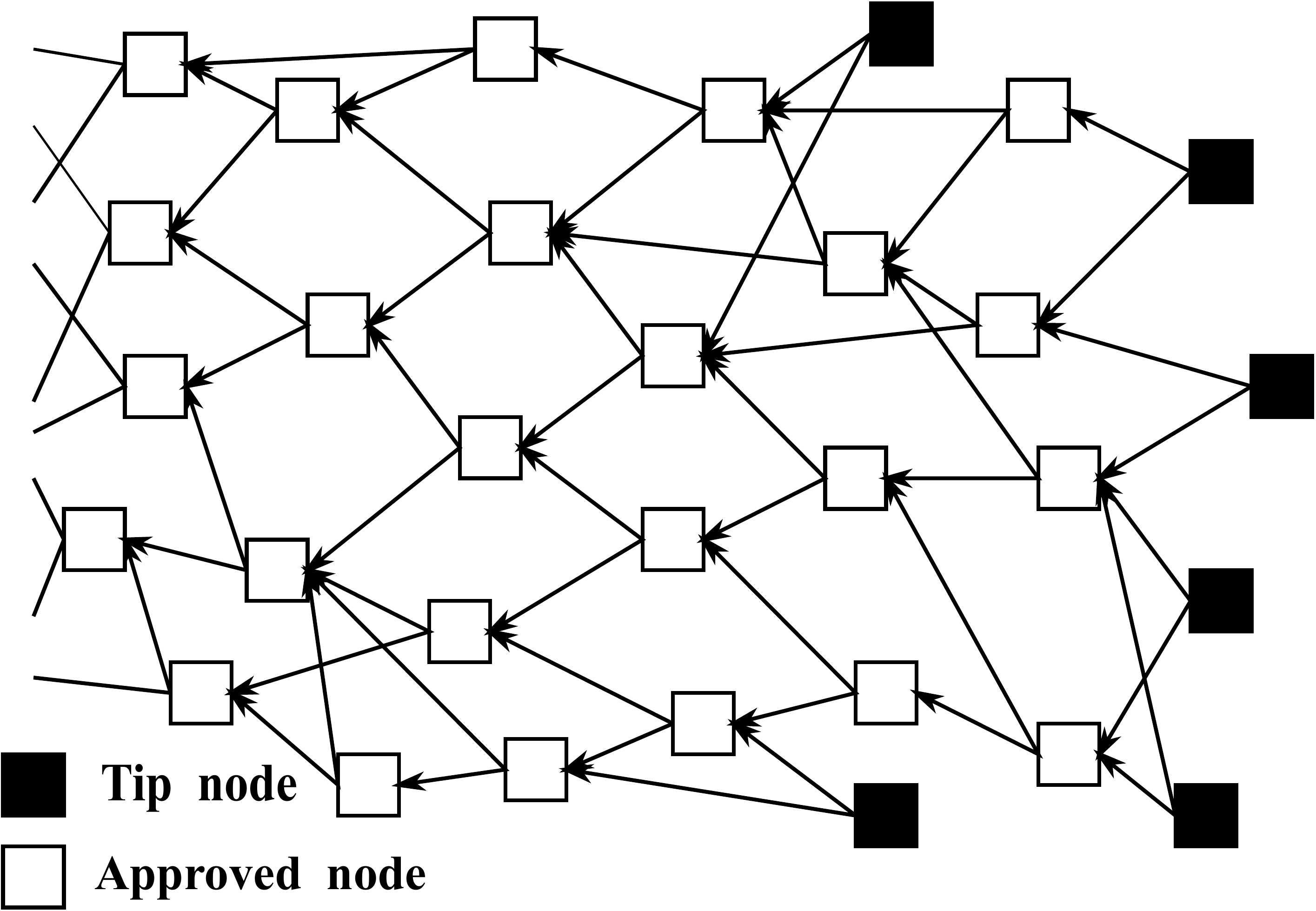}
 \caption{Tangle protocol}
  \label{tanglefig}
  \vspace{-5mm}
\end{figure}

Tangle is a new technology for distributed ledgers proposed by the cryptocurrency Iota. Tangle does not require a complicated, time consuming and computational-intensive consensus protocol. It also does not use blocks to store transactions. Each transaction is a unique block by itself which must approve two older transactions in order to be added to the ledger. Tangle uses Directed Acyclic Graph (DAG) in which each transaction is linked to two older transactions which are approved by it. After a transaction approves two older transactions, it is added to the ledger through proof of work.
The tangle framework is illustrated in Fig.~\ref{tanglefig}. Tip nodes, indicated by solid black squares, 
are the new transactions waiting for approval. The approved transactions are shown by the white squares. 
Note, each transaction is linked to two older transactions.


Due to the unique design of tangle, it is a fast, infinitely scalable framework which makes it well-suited for IoT networks. Furthermore, Tangle has no transaction fees which is desirable for an IoT network. In contrast to most of the blockchain implementations, tangle is immune to becoming obsolete with the advent of quantum computers because of its unique design. The main challenge about tangle is how to choose the two older transactions for approval. No rule is imposed by tangle on how to choose these two nodes which is very desirable for resource-constrained devices in an IoT network. However, the chosen transactions should not be the same or conflicting. To choose between conflicting transactions, tangle runs an algorithm called tip selection algorithm multiple times to figure out which of these two transactions are more plausible to be indirectly approved by the selected tip.


Unlike blockchain frameworks, tangle's design enables parallel transaction verification which eliminates the required wait time for mining previous blocks as in blockchain and provides the opportunity to verify more 
transactions in a shorter time. Although tangle is very promising and claims to overcome the existing barriers for decentralization of resource-constrained IoT networks~\cite{popov2016tangle}, 
it confronts a lot of implementation challenges, specifically for IoT applications. 
The current implementation of tangle, Iota, does not provide all the claimed goals of tangle. 
One of the challenges for applying tangle to IoT networks is the storage limitation. 
The resource constrained IoT devices are unable to store the entire tangle. Some solutions including automated snapshotting and a swarm client have been proposed to address this problem 
in Iota's development road map~\cite{roadmap}. Another problem with tangle is that whoever gains control over more than
one-third hash power of tangle can make it insecure and vulnerable. As a preventive measure, 
Iota runs a node called `coordinator' by amassing the hash power itself at one point. 
However, this can be perceived as centralization of tangle.

\vspace{-2mm}
\subsection{Comparisons of Different Consensus Methods}
The above mentioned consensus methods have been used in various blockchain implementations. 
The consensus method is the backbone of a blockchain implementation. Thus, most of the features and performance attributes of a blockchain implementation are contingent upon the method of consensus used~\cite{mingxiao2017review}. 

Researchers mostly propose new consensus protocols based on the cryptocurrency requirements. The most important features for a cryptocurrency are high throughput and security while for IoT networks, low latency is the most significant feature. In a practical IoT network, a transaction should be sent and finalized within a few milliseconds or less. In table~\ref{table-consensus}, all the mentioned consensus methods are compared to verify their applicability to IoT networks. Most appropriate consensus methods for IoT networks are denoted with {\begin{tikzpicture}
\draw (0,0) circle (0.12cm);
\clip (0,0) circle (0.12cm);
\fill[black] (0.12cm,0.12cm) rectangle (-0.12cm,-0.12cm);
\end{tikzpicture}}, partially apt consensus methods are denoted with {\begin{tikzpicture}
\draw (0,0) circle (0.12cm);
\clip (0,0) circle (0.12cm);
\fill[black] (0cm,0.12cm) rectangle (-0.12cm,-0.12cm);
\end{tikzpicture}}, and not applicable methods to IoT networks are denoted with {\begin{tikzpicture}
\draw (0,0) circle (0.12cm);
\clip (0,0) circle (0.12cm);
\fill[black] (0cm,0cm) rectangle (-0cm,-0cm);
\end{tikzpicture}}.

\section{Overview of Blockchain Implementations}
\label{sec-implementations}
In order to achieve a blockchain based IoT network, the most appropriate framework must be designed and
deployed. One option is to implement a new framework and use the preferred method of consensus which we believe best suits the IoT application requirements. The other option is to use 
an already defined framework. 
In this section, we review some of the available frameworks for implementing a blockchain system and 
their practical applicability towards an IoT application. 
We survey different aspects of these implementations including their method of consensus, their applications, being permissioned or permissionless, being private or public. All these features affect the characteristics of a blockchain framework along with its scalability, performance and availability. Scalability refers to the size of a blockchain network and number of users it can support. Performance refers to the latency and throughput which are 
critical for IoT networks. Latency is the amount of time it takes for the nodes reach to a consensus. 
Low latency can be achieved by compromising the decentralization of the blockchain framework. 
Throughput is the number of transactions that can be processed per time unit. 
Availability refers to the accessibility of the nodes to a copy of the distributed ledger~\cite{zheng2016blockchain}.

The various private blockchain implementations are presented in Section~\ref{implement-private}. 
One of the well-known frameworks for implementing a private blockchain based system is the 
Hyperledger framework. Hyperledger is a collaborative project started by the Linux Foundation and developed by many companies including IBM, Intel, Cisco, Hitachi and so on. 
It contains several projects including fabric, sawtooth, indy, iroha and burrow. IBM blockchain for IoT is another private blockchain implementation. 
As part of the public blockchain based system, we discuss the well-known ones: ethereum and bitcoin
in Section~\ref{implement-public}
We also survey the implementations for alternative technologies for blockchain including corda and iota in Section~\ref{implement-alternative}. 
We compare the most promising frameworks for IoT applications in Section~\ref{implement-comparison}.

\vspace*{-1pt}
\subsection{Private Blockchains}
\label{implement-private}
\vspace*{-1pt}

\noindent
{\underline{\em Hyperledger Fabric:}}
Hyperledger Fabric is a permissioned implementation which is widely used by enterprises. It uses a
pluggable method of consensus which is defined based on specific application requirements. 
Its most common method of consensus is PBFT. Unlike bitcoin  and ethereum which are public blockchains, 
this private blockchain framework attains consensus within hundreds of milliseconds~\cite{red2017practical}. 
Such low latency is crucial for building blockchain based IoT networks.
Being permissioned, the blockchain is controlled by a specific organization which allows specific nodes to join the blockchain, access to the database and participate in the consensus protocol. This framework supports {\tt chaincode} designed in the {\tt Go} language which is a special version of smart contracts~\cite{cachin2016architecture}.   
 Smart contract is a self-executing software written in a programming language that allows users to program their own scripts for transferring financial assets, products, or services between different parties without a middleman~\cite{buterin2014next}.
 Although low latency of this implementation is a noticeable advantage for IoT networks, it remains a private blockchain and therefore 
lacks the beneficial features of public blockchains such as being totally distributed with highly secure and immutable data storage.
Furthermore, the network overhead for this framework significantly increases with increase in the number of nodes which causes the number of messages communicated in the PBFT protocol to increase. This prevents hyperledger fabric to be used in large scale applications similar to public blockchains.

Hyperledger fabric consists of several entities that maintain the blockchain: peers, orderer, chaincode and chaincode policy. Peers are responsible for validating and performing the requests by the nodes within the network. The validation of new blocks is also the responsibility of the peers. 
The orderer receives all the valid invoke requests called transactions from nodes, creates a
block from of these transactions, sends them to peers to be verified, 
and adds to their copies of the ledger. These invoke requests are first verified by one or more of the corresponding peers which are dedicated according to the defined policy in the chaincode. Thereafter, these requests are sent to the orderer for further process.
Different kinds of messages can be defined on the chaincode by which nodes propagate different requests to peers. The most used messages for a chaincode are query and invoke. Query is getting to know the current state of the ledger which is like asking the current value of a variable assigned to a node.
Invoke is requesting a change in the ledger which is like changing the value
of a variable assigned to a node~\cite{roy2019cache}.

\vspace*{5pt}
\noindent
{\underline{\em Hyperledger Sawtooth:}}
Proposed by Intel, Hyperledger Sawtooth is a modular platform for implementation of distributed ledgers for storing digital records aptly designed for enterprise usage. It uses proof of elapsed time using Intel's Software Guard Extensions (SGX) as a trusted execution environment for achieving consensus. This platform allows large scale implementation of both permissioned and permissionless ledgers, 
and has features such as  live data stream, hardware security and enterprise-grade customer load which make it suitable for IoT devices~\cite{dhillon2017hyperledger}. However, this framework is not yet implemented in large scale and fully tested for its performance capabilities. 
It is also not recommended by Intel to be used for security sensitive applications due to its lack of security mechanisms.


\vspace*{5pt}
\noindent
{\underline{\em Hyperledger Indy:}}
Hyperledger Indy is a distributed ledger framework designed specifically for decentralized identities to prevent digital identity breaches on the Internet. It utilizes Zero-Knowledge Proofs (ZKP) to avoid inessential revelation of identity features. It is a permissioned blockchain but with global public access to its features. For validating new blocks, it uses an approach called Plenum in which a set of validator nodes run a modified, redundant Byzantine fault tolerant protocol~\cite{hyperledger}. There is no striking aspect of 
Hyperledger Indy that makes it attractive for an IoT network.

\vspace*{5pt}
\noindent
{\underline{\em Hyperledger Iroha:}}
Hyperledger Iroha is a simple implementation that focuses on mobile application development of blockchain technology~\cite{dhillon2017hyperledger}. It is developed in {\tt C++} and uses a new method of consensus called Sumeragi which is a chain-based Byzantine fault tolerant consensus algorithm. In this framework, data storage and synchronization are performed off-device~\cite{hyperledger}. In Iroha, every participant is not permitted to access to entire data history. The participants can gain access to data through query by getting permission and being authenticated~\cite{saraf2018blockchain}.

\vspace*{5pt}
\noindent
{\underline{\em Hyperledger Burrow:}}
Hyperledger Burrow is a permissioned blockchain that uses Tendermint consensus method for achieving consensus~\cite{wang2019blockchain}. It was first designed by a company called Monax. Its design is based on Ethereum Virtual Machine (EVM). It is a general-purpose smart contract machine for cross-industry applications and is not an optimal framework for a single industry~\cite{hyperledger}. The main drawback of this implementation is that if one-third or more number of validators become offline, the network may halt~\cite{saraf2018blockchain}.

\vspace*{5pt}
\noindent
{\underline{\em IBM Watson IoT:}}
Based on Hyperledger Fabric project, IBM Watson IoT is private blockchain framework for IoT networks proposed by IBM~\cite{IBM}. Though very appropriate for
small scale IoT networks, scalability remains the main drawback. Each participant has the full solution functionality up to 10 routes (IoT to Blockchain connections), with up to 1 transaction per second per route. 

\subsection{Public Blockchains}
\label{implement-public}

\noindent
{\underline{\em Bitcoin:}}
The well know cryptocurrency, bitcoin, utilizes a permissionless public blockchain framework. 
Being permissionless, it allows any node to participate in the consensus protocol and mine blocks
without any permission. It uses proof of work as the method for consensus which has a high latency about 10 minutes, making it ineffective for IoT networks. However, it is worth exploring if it can still be
used with eased proof of work to reach a consensus in a short time. Alternatively, this framework
can be used in in combination with other methods. Another significant challenge in applying bitcoin blockchain to IoT networks is its scalability. Average transaction size for bitcoin is between 400 Bytes and 600 Bytes. Considering the size of each bitcoin's block which is 1 MB and the average time to solve each block which is 10 minutes, current bitcoin architecture allows 4 transactions per second which is not acceptable for IoT networks. 

\vspace*{5pt}
\noindent
{\underline{\em Ethereum:}}
Ethereum is a permissionless public blockchain framework developed using solidity 
which is a contract-oriented, high-level language for implementing smart contracts~\cite{valenta2017comparison}. All the nodes are required to participate in the 
consensus process. It was primarily designed using a derivative of proof of work known as Ethash. This method is significantly less computational-intensive than the original proof of work because of using Directed Acyclic Graph (DAG). This blockchain can be customized and adopted for a variety of applications because of its intrinsic characteristics that enable smart contracts~\cite{red2017practical}. 
There are plans for migrating Ethereum to Casper~\cite{zheng2016blockchain}. Its block generation process takes between 10 to 20 seconds which is much less than bitcoin's latency but still is not practical for an IoT network implementation~\cite{red2017practical}. 
It should be noted that some private and permissioned blockchain frameworks are also designed based on the Ethereum blockchain.

\begin{table*}[!h]
\scriptsize
\caption{Comparisons of various blockchain implementations}
\centering
\begin{tabular}{|l|c|c|c|c|c|c|}
\hline
  \multirow{2}{*} {\backslashbox{\textbf{Features}}{\textbf{Implement.}}} & \textbf{Hyperledger} & \textbf{Hyperledger} & \multirow{2}{*} {\textbf{Bitcoin}} & \multirow{2}{*} {\textbf{Ethereum}} & \multirow{2}{*} {\textbf{Corda}} & \multirow{2}{*} {\textbf{Iota}}\\
& \textbf{Fabric} & \textbf{Sawtooth} &  &  &  & \\
\hline
\multirow{2}{*} {\textbf{Consensus method}} & Pluggable & Proof of & Proof of & Ethash (PoW) & Pluggable & \multirow{2}{*} {Tangle} \\
 & (PBFT generally) & elapsed time & Work & Casper (PoS) & (Raft generally) & \\
\hline
\multirow{2}{*} {Accessibility} & \multirow{2}{*} {Private} & \multirow{2}{*} {Private} & \multirow{2}{*} {Public} & \multirow{2}{*} {Public} & \multirow{2}{*} {Private} & \multirow{2}{*} {Public}\\
&  &  &  &  &  & \\
\hline
\multirow{2}{*}{Mode of operation} & \multirow{2}{*}{Permissioned} & Permissioned or & \multirow{2}{*}{Permissionless} & Permissioned or & \multirow{2}{*}{Permissioned} & \multirow{2}{*}{Permissionless} \\
 &  &  Permissionless &  & Permissionless &  & \\
\hline
\multirow{2}{*} {Decentralization} & \multirow{2}{*} {Partially} & \multirow{2}{*} {Partially} & \multirow{2}{*} {Yes} & \multirow{2}{*} {Yes} & \multirow{2}{*} {Partially} & \multirow{2}{*} {Partially}\\
&  &  &  &  &  & \\
\hline
\multirow{2}{*} {Compute-intensive} & \multirow{2}{*} {No} & \multirow{2}{*} {No} & \multirow{2}{*} {Yes} & \multirow{2}{*} {Partially} & \multirow{2}{*} {No} & \multirow{2}{*} {No} \\
&  &  &  &  &  &  \\
\hline
\multirow{2}{*} {Network-intensive} & \multirow{2}{*} {Yes} & \multirow{2}{*} {No} & \multirow{2}{*} {No} & \multirow{2}{*} {No} & \multirow{2}{*} {No} & \multirow{2}{*} {No}\\
&  &  &  &  &  & \\
\hline
\multirow{2}{*} {Scalability} & \multirow{2}{*} {Low} & \multirow{2}{*} {High} & \multirow{2}{*} {High} & \multirow{2}{*} {High} & \multirow{2}{*} {Partially} & \multirow{2}{*} {High}\\
&  &  &  &  &  & \\
\hline
\multirow{2}{*} {Throughput} & \multirow{2}{*} {High} & \multirow{2}{*} {High} & \multirow{2}{*} {Very low} & \multirow{2}{*} {Low} & \multirow{2}{*} {High} & \multirow{2}{*} {High}\\
&  &  &  &  &  & \\
\hline
\multirow{2}{*} {Latency} & \multirow{2}{*} {100 ms} & \multirow{2}{*} {Very Low} & \multirow{2}{*} {10 Minutes} & \multirow{2}{*} {12 Seconds} & Very Low & \multirow{2}{*} {10 ms}\\
&  &  &  &  & Not Measured & \\
\hline
\multirow{2}{*} {Immutability} & \multirow{2}{*} {Low} & \multirow{2}{*} {Low} & \multirow{2}{*} {High} & \multirow{2}{*} {High} & \multirow{2}{*} {High} & \multirow{2}{*} {High}\\
&  &  &  &  &  & \\
\hline
\multirow{2}{*} {Adversary tolerance} & 33.33\% & \multirow{2}{*} {Unverified} & \textless25\% & \textless51\% & \multirow{2}{*} {unverified} & 33.33\%\\
& Faulty Replicas &  & Computing Power & Stakes &  & Computing Power\\
\hline
\multirow{2}{*} {Privacy} & \multirow{2}{*} {High} & \multirow{2}{*} {High} & \multirow{2}{*} {Low} & \multirow{2}{*} {Low} & \multirow{2}{*} {High} & \multirow{2}{*} {Low}\\
&  &  &  &  &  & \\
\hline
\multirow{2}{*} {Smart contract} & \multirow{2}{*} {Yes} & \multirow{2}{*} {Yes} & \multirow{2}{*} {Limited} & \multirow{2}{*} {Yes} & \multirow{2}{*} {Yes} & \multirow{2}{*} {No}\\
&  &  &  &  &  & \\
\hline
\multirow{2}{*}{Currency} & None but & None but & \multirow{2}{*}{Bitcoin (BTC)} & Ether (ETH), & \multirow{2}{*}{None} & \multirow{2}{*}{Iota}\\
& Tokens possible & Tokens possible &  & Tokens possible &  & \\
\hline
\end{tabular}
\label{implementation-comparison}
\end{table*}

\subsection{Blockchain Alternatives}
\label{implement-alternative}

\noindent
{\underline{\em Corda:}}
Corda is a permissioned decentralized ledger framework that uses pluggable method of consensus. In this implementation, specific nodes called notary nodes are responsible for the consensus protocol. Due to the requirement of trusting the notaries for consensus, it is partially decentralized. It cannot be considered as a blockchain framework since it uses a different architecture. This framework is specifically designed for financial applications and not very suitable for resource constrained IoT networks~\cite{valenta2017comparison}.

\vspace*{5pt}
\noindent
{\underline{\em Iota:}}
Iota is a distributed ledger that uses Directed Acyclic Graph (DAG) instead of a blockchain. This protocol is called Tangle which was discussed in section~\ref{sec-tangle}. 
 Its fast speed of transactions is very desirable for IoT applications and is the 
 first cryptocurrency which has been specifically designed for IoT applications. 
This design minimizes the transaction time and network overhead at the cost of relaxing security with some possible attack scenarios~\cite{red2017practical}.
Iota does not have any transaction fee which enables micro transactions. Unlike other cryptocurrencies that use a number of confirmations as an indicator for the reliability of a transaction, Iota uses cumulative weight which is defined for each transaction based on the number of consecutive linked transactions to it~\cite{popov2016tangle}. In Iota, all the tokens have been made available from the first day and they were released by the first node called the genesis transaction.

\subsection{Comparisons of Various Implementations}
\label{implement-comparison}
In order to assess which implementation addresses most of the limitations of blockchain and is applicable to IoT networks, 
we compare six implementations in Table~\ref{implementation-comparison}. 
A desirable blockchain implementation for IoT networks should have the following features: decentralized, not compute-intensive, not network-intensive, high scalability, high throughput, very low latency and preserving privacy. 
We do not compare Indy, Iroha, Burrow and IBM Watson since they have been primarily designed for specific purposes and are not
necessarily suitable for IoT networks. Moreover, these frameworks have not yet been fully developed and implemented. 

The mentioned features in Table~\ref{implementation-comparison} for Hyperledger Sawtooth are claimed in theory and they are not yet fully tested in a large scale implementation under different circumstances. In addition, Hyperledger Sawtooth is still under experiment and not recommended by Intel to be used for security sensitive applications. High throughput and low latency of Hyperledger Fabric is in small-scale implementations and for large-scale networks they will deteriorate. The latency values in Table~\ref{implementation-comparison} are average values. Bitcoin and Ethereum have high scalability only with very limited network throughput. In Corda framework, a new transaction is dependent on many other transactions and if a node has not seen older transactions formerly will result in a significant high latency which pose some restrictions on scalability. However, generally, its latency is very low. Furthermore, it is claimed in Corda's white paper that it is secure against adversarial attacks which is not tested and implemented yet. Low tolerance of Iota's framework against adversarial activities is solved by using a special node called coordinator.

\section{Open Research Challenges}
\label{research-section}
Considering the unique features of blockchain, it can be applied to a number of domains including but not limited to IoT networks, healthcare, data storage, inventory tracking and finance. The primary challenge is how to adapt the blockchain technology to suit the specific application needs. As every application poses different requirements, a new or a customized implementation of blockchain is needed. In the IoT domain, the major research challenges that ought to be addressed are:
\begin{itemize}
\item {\em Enhancing scalability:} Scalability refers to the size of a blockchain network and the number of users it can support. In a practical IoT network, a large number of devices need to communicate through the network. Enhancing scalability using the current implementations, jeopardize throughput and latency.
\item {\em Guaranteeing security:} There are lots of malicious activities that target IoT networks. A practical IoT network postulates immunity to all of the plausible attacks. Current state of the art security methods rely on sophisticated computations. Therefore, securing a network with resource-constrained devices unable to perform heavy-duty computations is a challenge.
\item {\em Protecting data privacy:} The communicated data between different entities are confidential. Therefore, preserving privacy in communication links is a necessity. Making a blockchain-based IoT network more private, usually endanger the basic paradigm of decentralization in blockchains.
\item {\em Increasing throughput:} In an IoT network, a large number of devices are required to simultaneously communicate with each other which necessitates a network with high throughput. Increasing throughput in current implementations usually reduces scalability which is not desirable.
\item {\em Reducing latency:} In a practical IoT network, different devices need to communicate with each other in real-time. Therefore, the latency should be very low. Reducing latency in current implementations usually compromises scalability, which is not acceptable.
\item {\em Reducing computational requirements:} IoT devices are mostly resource-constrained. Current state of the art security protocols mostly rely on sophisticated cryptographic computations which is a burden for resource-constrained devices. 
\item {\em Overcoming storage limitations:} Resource-constrained IoT devices cannot store huge amounts of data which is usually a requirement for blockchain based networks. This is because, several nodes or sometimes all the nodes are typically required to have a copy of the blockchain.
\end{itemize}

In order to successfully apply blockchain to IoT networks, practically feasible solutions are required that are scalable to large networks and yield high throughput with low latency. In addition, the implementations need to be highly secure in order to defend against possible attacks and be tamper-resistant. Besides, the implementations should be compatible with resource-constrained IoT devices that have limited computational capability and restricted storage capacity. Each of the discussed consensus methods and implementations addresses several of the above mentioned issues. However, there remains the need for implementations that address all the mentioned challenges.

\section{Conclusions}
\label{sec-5}
In this article, we discussed the possibilities of using blockchain for securing and 
assuring data integrity in IoT networks. In particular, we focused on the currently used
consensus methods and their practical applicability for resource constrained IoT devices and networks. 
We discussed the pros and cons of current consensus methods used in blockchain implementations.
We also discussed how private blockchains and tangle can be a better alternative
to public blockchains for IoT networks. 

Among the discussed implementations of blockchain, Hyperledger Fabric, Sawtooth, Iota, 
and Ethereum appear more promising for IoT networks and applications since they have addressed some of the existing limitations of blockchain. 
Each of these implementations have addressed some of the limitations including throughput, latency, computational overhead, network overhead, scalability and privacy. However, none of them have been successful in addressing all the limitations to an acceptable degree.
We believe that in order to realize a blockchain based IoT network on a large scale and 
with low latency, there ought to be either a hybrid framework which combines two or more of the already existing frameworks or an existing framework that have a modified method of consensus.


\begin{thebibliography}{11}

 \bibitem{bhattacharjee2017preserving}
 Bhattacharjee, Shameek, Mehrdad Salimitari, Mainak Chatterjee, Kevin Kwiat, and Charles Kamhoua. "Preserving Data Integrity in IoT Networks Under Opportunistic Data Manipulation." In Dependable, Autonomic and Secure Computing (DASC), 2017 IEEE 15th Intl, pp. 446-453.

 \bibitem{debus2017consensus}
 Debus, Julian. "Consensus methods in blockchain systems." Frankfurt School of Finance \& Management, Blockchain Center, Tech. Rep (2017).

 \bibitem{zheng2016blockchain}
 Zheng, Zibin, Shaoan Xie, Hong-Ning Dai, and Huaimin Wang. "Blockchain challenges and opportunities: A survey." Work Pap.–2016 (2016).

 \bibitem{sehgal2012management}
 Sehgal, Anuj, Vladislav Perelman, Siarhei Kuryla, and Jurgen Schonwalder. "Management of resource constrained devices in the internet of things." IEEE Communications Magazine 50, no. 12 (2012).

 \bibitem{popov2016tangle}
 Popov, Serguei. "The tangle." cit. on (2016): 131.

 \bibitem{salimitari2017profit}
 Salimitari, Mehrdad, Mainak Chatterjee, Murat Yuksel, and Eduardo Pasiliao. "Profit Maximization for Bitcoin Pool Mining: A Prospect Theoretic Approach." In Collaboration and Internet Computing (CIC), 2017 IEEE 3rd International Conference on, pp. 267-274. IEEE, 2017.

 \bibitem{nakamoto2008bitcoin}
 Nakamoto, Satoshi. "Bitcoin: A peer-to-peer electronic cash system." (2008).

 \bibitem{Eyal:2018}
 Eyal, Ittay, and Emin Gün Sirer. "Majority is not enough: Bitcoin mining is vulnerable." Communications of the ACM 61, no. 7 (2018): 95-102.

 \bibitem{banafa2017iot}
 Banafa, Ahmed. "IoT and Blockchain Convergence: Benefits and Challenges." IEEE Internet of Things (2017).
 


 \bibitem{wang2015processing}
 Wang, Lizhe, and Rajiv Ranjan. "Processing distributed internet of things data in clouds." IEEE Cloud Computing 2, no. 1 (2015): 76-80.
 
 \bibitem{fernandez2018review}
 Fernández-Caramés, Tiago M., and Paula Fraga-Lamas. "A Review on the Use of Blockchain for the Internet of Things." IEEE Access 6 (2018): 32979-33001.
 
 \bibitem{triantafillou2003nanopeer}
 Triantafillou, Peter, Nikos Ntarmos, S. Nikoletseas, and P. Spirakis. "NanoPeer networks and P2P worlds." In Proceedings Third International Conference on Peer-to-Peer Computing (P2P2003), pp. 40-46. IEEE, 2003.
 
 \bibitem{ali2004csn}
 Ali, Muneeb, and Zartash Afzal Uzmi. "CSN: A network protocol for serving dynamic queries in large-scale wireless sensor networks." In Proceedings. Second Annual Conference on Communication Networks and Services Research, 2004., pp. 165-174. IEEE, 2004.
 
 \bibitem{krco2005p2p}
 Krco, Srdjan, David Cleary, and Daryl Parker. "P2P mobile sensor networks." In Proceedings of the 38th Annual Hawaii International Conference on System Sciences, pp. 324c-324c. IEEE, 2005.
 
 \bibitem{kshetri2017can}
 Kshetri, Nir. "Can blockchain strengthen the internet of things?." IT professional 19, no. 4 (2017): 68-72.
 
 \bibitem{anirudh2017use}
 Anirudh, M., S. Arul Thileeban, and Daniel Jeswin Nallathambi. "Use of honeypots for mitigating DoS attacks targeted on IoT networks." In 2017 International Conference on Computer, Communication and Signal Processing (ICCCSP), pp. 1-4. IEEE, 2017.
 
 \bibitem{xu2016security}
 Xu, Qian, Pinyi Ren, Houbing Song, and Qinghe Du. "Security enhancement for IoT communications exposed to eavesdroppers with uncertain locations." IEEE Access 4 (2016): 2840-2853.
 
 \bibitem{li2017iot}
 Li, Xin, Huazhe Wang, Ye Yu, and Chen Qian. "An IoT data communication framework for authenticity and integrity." In 2017 IEEE/ACM Second International Conference on Internet-of-Things Design and Implementation (IoTDI), pp. 159-170. IEEE, 2017.
 
 \bibitem{yu2017recursive}
 Yu, Tianqi, Xianbin Wang, and Abdallah Shami. "Recursive principal component analysis-based data outlier detection and sensor data aggregation in iot systems." IEEE Internet of Things Journal 4, no. 6 (2017): 2207-2216.
 
 \bibitem{wang2019survey}
 Wang, Xu, Xuan Zha, Wei Ni, Ren Ping Liu, Y. Jay Guo, Xinxin Niu, and Kangfeng Zheng. "Survey on blockchain for Internet of Things." Computer Communications (2019).
 
 \bibitem{lin2017survey}
 Lin, Jie, Wei Yu, Nan Zhang, Xinyu Yang, Hanlin Zhang, and Wei Zhao. "A survey on internet of things: Architecture, enabling technologies, security and privacy, and applications." IEEE Internet of Things Journal 4, no. 5 (2017): 1125-1142.
 
 \bibitem{khan2018iot}
 Khan, Minhaj Ahmad, and Khaled Salah. "IoT security: Review, blockchain solutions, and open challenges." Future Generation Computer Systems 82 (2018): 395-411.
 
 \bibitem{lin2017using}
 Lin, Jun, Zhiqi Shen, and Chunyan Miao. "Using blockchain technology to build trust in sharing LoRaWAN IoT." In Proceedings of the 2nd International Conference on Crowd Science and Engineering, pp. 38-43. ACM, 2017.
 
 \bibitem{novo2018blockchain}
 Novo, Oscar. "Blockchain meets IoT: An architecture for scalable access management in IoT." IEEE Internet of Things Journal 5, no. 2 (2018): 1184-1195.
 
 \bibitem{hyperledger}
 "Hyperledger", \url{https://www.hyperledger.org}, Accessed: 2018-04-16
 
 \bibitem{bitshare}
 "Bitshares", \url{https://bitshares.org}, Accessed: 2019-05-23
 
 \bibitem{larimer2014delegated}
 Larimer, Daniel. "Delegated proof-of-stake (dpos)." Bitshare whitepaper (2014).
 
  \bibitem{lease}
 "Leased Proof of Stake", \url{https://docs.wavesplatform.com/platform-features/leased-proof-of-stake-lpos.html}, Accessed: 2018-04-25

 \bibitem{importance}
 "Proof of Importance",
 \url{https://nem.io/technology/}, Accessed: 2018-04-05
 
 \bibitem{buterin2017casper}
 Buterin, Vitalik, and Virgil Griffith. "Casper the friendly finality gadget." arXiv preprint arXiv:1710.09437 (2017).
 
 
 \bibitem{sompolinsky2015secure}
 Sompolinsky, Yonatan, and Aviv Zohar. "Secure high-rate transaction processing in bitcoin." In International Conference on Financial Cryptography and Data Security, pp. 507-527. Springer, 2015.
 
 \bibitem{zamfir2017casper}
 Zamfir, Vlad. "Casper the friendly ghost: A correct by construction blockchain consensus protocol." (2017).
 
 \bibitem{mazieres2015stellar}
 Mazieres, David. "The stellar consensus protocol: A federated model for internet-level consensus." Stellar Development Foundation (2015).
 
 \bibitem{dinh2018untangling}
 Dinh, Tien Tuan Anh, Rui Liu, Meihui Zhang, Gang Chen, Beng Chin Ooi, and Ji Wang. "Untangling blockchain: A data processing view of blockchain systems." IEEE Transactions on Knowledge and Data Engineering 30, no. 7 (2018): 1366-1385.
 
 \bibitem{sankar2017survey}
 Sankar, Lakshmi Siva, M. Sindhu, and M. Sethumadhavan. "Survey of consensus protocols on blockchain applications." In 2017 4th International Conference on Advanced Computing and Communication Systems (ICACCS), pp. 1-5. IEEE, 2017.
 
 \bibitem{schwartz2014ripple}
 Schwartz, David, Noah Youngs, and Arthur Britto. "The ripple protocol consensus algorithm." Ripple Labs Inc White Paper 5 (2014).
 
 \bibitem{zheng2017overview}
 Zheng, Zibin, Shaoan Xie, Hongning Dai, Xiangping Chen, and Huaimin Wang. "An overview of blockchain technology: Architecture, consensus, and future trends." In 2017 IEEE International Congress on Big Data (BigData Congress), pp. 557-564. IEEE, 2017.
 
 \bibitem{chalaemwongwan2018state}
 Chalaemwongwan, Nutthakorn, and Werasak Kurutach. "State of the art and challenges facing consensus protocols on blockchain." In 2018 International Conference on Information Networking (ICOIN), pp. 957-962. IEEE, 2018.
 
 \bibitem{dib2018consortium}
 Dib, Omar, Kei-Leo Brousmiche, Antoine Durand, Eric Thea, and E. Ben Hamida. "Consortium blockchains: Overview, applications and challenges." International Journal On Advances in Telecommunications 11, no. 1\&2 (2018).
 
 \bibitem{Tendermint}
 "Tendermint", \url{https://tendermint.com}, Accessed: 2019-03-07
 
 \bibitem{kwon2014tendermint}
 Kwon, Jae. "Tendermint: Consensus without mining." Retrieved May 18 (2014): 2017.
 
 \bibitem{bach2018comparative}
 Bach, L. M., Branko Mihaljevic, and Mario Zagar. "Comparative analysis of blockchain consensus algorithms." In 2018 41st International Convention on Information and Communication Technology, Electronics and Microelectronics (MIPRO), pp. 1545-1550. IEEE, 2018.
 
 \bibitem{syta2016keeping}
  Syta, Ewa, Iulia Tamas, Dylan Visher, David Isaac Wolinsky, Philipp Jovanovic, Linus Gasser, Nicolas Gailly, Ismail Khoffi, and Bryan Ford. "Keeping authorities" honest or bust" with decentralized witness cosigning." In 2016 IEEE Symposium on Security and Privacy (SP), pp. 526-545. Ieee, 2016.
  
 \bibitem{kogias2016enhancing}
 Kogias, Eleftherios Kokoris, Philipp Jovanovic, Nicolas Gailly, Ismail Khoffi, Linus Gasser, and Bryan Ford. "Enhancing bitcoin security and performance with strong consistency via collective signing." In 25th {USENIX} Security Symposium, pp. 279-296. 2016.
   
 \bibitem{micali1999verifiable}
 Micali, Silvio, Michael Rabin, and Salil Vadhan. "Verifiable random functions." In 40th Annual Symposium on Foundations of Computer Science (Cat. No. 99CB37039), pp. 120-130. IEEE, 1999.  

 \bibitem{Algorand}
 "Algorand", \url{https://www.algorand.com}, Accessed: 2019-05-07
 
 \bibitem{gilad2017algorand}
 Gilad, Yossi, Rotem Hemo, Silvio Micali, Georgios Vlachos, and Nickolai Zeldovich. "Algorand: Scaling byzantine agreements for cryptocurrencies." In Proceedings of the 26th Symposium on Operating Systems Principles, pp. 51-68. ACM, 2017.
     
 \bibitem{zamani2018rapidchain}
  Zamani, Mahdi, Mahnush Movahedi, and Mariana Raykova. "RapidChain: Scaling blockchain via full sharding." In Proceedings of the 2018 ACM SIGSAC Conference on Computer and Communications Security, pp. 931-948. ACM, 2018.
  
 \bibitem{hanke2018dfinity}
 Hanke, Timo, Mahnush Movahedi, and Dominic Williams. "Dfinity technology overview series, consensus system." arXiv preprint arXiv:1805.04548 (2018).
 
 \bibitem{danezis2015centrally}
 Danezis, George, and Sarah Meiklejohn. "Centrally banked cryptocurrencies." arXiv preprint arXiv:1505.06895 (2015).
 
 \bibitem{luu2016secure}
 Luu, Loi, Viswesh Narayanan, Chaodong Zheng, Kunal Baweja, Seth Gilbert, and Prateek Saxena. "A secure sharding protocol for open blockchains." In Proceedings of the 2016 ACM SIGSAC Conference on Computer and Communications Security, pp. 17-30. ACM, 2016.
 
 \bibitem{kokoris2018omniledger}
 Kokoris-Kogias, Eleftherios, Philipp Jovanovic, Linus Gasser, Nicolas Gailly, Ewa Syta, and Bryan Ford. "Omniledger: A secure, scale-out, decentralized ledger via sharding." In 2018 IEEE Symposium on Security and Privacy (SP), pp. 583-598. IEEE, 2018.
 
 \bibitem{mingxiao2017review}
 Mingxiao, Du, Ma Xiaofeng, Zhang Zhe, Wang Xiangwei, and Chen Qijun. "A review on consensus algorithm of blockchain." In 2017 IEEE International Conference on Systems, Man, and Cybernetics (SMC), pp. 2567-2572. IEEE, 2017.
 
 \bibitem{ongaro2014search}
 Ongaro, Diego, and John Ousterhout. "In search of an understandable consensus algorithm." In 2014 {USENIX} Annual Technical Conference ({USENIX}{ATC} 14), pp. 305-319. 2014.
 
 \bibitem{xu2017taxonomy}
 Xu, Xiwei, Ingo Weber, Mark Staples, Liming Zhu, Jan Bosch, Len Bass, Cesare Pautasso, and Paul Rimba. "A taxonomy of blockchain-based systems for architecture design." In 2017 IEEE International Conference on Software Architecture (ICSA), pp. 243-252. IEEE, 2017.
 
 \bibitem{bano2017consensus}
 Bano, Shehar, Alberto Sonnino, Mustafa Al-Bassam, Sarah Azouvi, Patrick McCorry, Sarah Meiklejohn, and George Danezis. "Consensus in the age of blockchains." arXiv preprint arXiv:1711.03936 (2017).
 
 \bibitem{roadmap}
  "Iota Development Roadmap", \url{https://blog.iota.org},
  note = Accessed: 2018-04-14
  
  \bibitem{red2017practical}
 Red, Val A. "Practical comparison of distributed ledger technologies for IoT." In Disruptive Technologies in Sensors and Sensor Systems, vol. 10206, International Society for Optics and Photonics, 2017.
 
 \bibitem{cachin2016architecture}
 Cachin, Christian. "Architecture of the hyperledger blockchain fabric." In Workshop on distributed cryptocurrencies and consensus ledgers, vol. 310. 2016.
      
 \bibitem{buterin2014next}
 Buterin, Vitalik. "A next-generation smart contract and decentralized application platform." white paper (2014).
 
 \bibitem{roy2019cache}
 Roy, Swapnoneel, Faustina J. Anto Morais, Mehrdad Salimitari, and Mainak Chatterjee. "Cache attacks on blockchain based information centric networks: an experimental evaluation." In Proceedings of the 20th International Conference on Distributed Computing and Networking, pp. 134-142. ACM, 2019.
 
 \bibitem{dhillon2017hyperledger}
 Dhillon, Vikram, David Metcalf, and Max Hooper. "The Hyperledger Project." In Blockchain Enabled Applications, pp. 139-149. Apress, Berkeley, CA, 2017.
 
 \bibitem{saraf2018blockchain}
 Saraf, Chinmay, and Siddharth Sabadra. "Blockchain platforms: A compendium." In 2018 IEEE International Conference on Innovative Research and Development (ICIRD), pp. 1-6. IEEE, 2018.
 
 \bibitem{wang2019blockchain}
 Wang, Shuai, Liwei Ouyang, Yong Yuan, Xiaochun Ni, Xuan Han, and Fei-Yue Wang. "Blockchain-Enabled Smart Contracts: Architecture, Applications, and Future Trends." IEEE Transactions on Systems, Man, and Cybernetics: Systems (2019).
 
 \bibitem{IBM}
 "IBM Watson IoT", \url{https://www.ibm.com/internet-of-things/spotlight/blockchain}, Accessed: 2018-04-25

 \bibitem{valenta2017comparison}
 Valenta, Martin, and Philipp Sandner. Comparison of Ethereum, Hyperledger Fabric and Corda. FSBC Working Paper, 2017.

 \end{thebibliography}
\end{document}